\RequirePackage{fix-cm}

\documentclass[twocolumn]{svjour3}

\smartqed  

\usepackage{graphicx}
\usepackage{tikz}
\usetikzlibrary{shapes,arrows,arrows.meta,calc,fit}
\usepackage{pgfplots}
\usepackage{subcaption}
\usepackage{amssymb}
\usepackage{amsmath}
\usepackage{stmaryrd}
\usepackage{url}

\DeclareMathAlphabet{\mathpzc}{OT1}{pzc}{m}{it}

\journalname{ArXiv}

\definecolor{EDIT_COLOR}{rgb}{0, 0, 0}

\begin{document}

\title{ParticLS: Object-oriented software for discrete element methods and peridynamics}


\author{Andrew D. Davis         \and
        Brendan A. West         \and
        Nathanael J. Frisch     \and
        Devin T. O'Connor       \and
        Matthew D. Parno
}

\institute{Andrew D. Davis \at
              Cold Regions Research and Engineering Laboratory, 72 Lyme Rd, Hanover, NH 03755 \\
              NYU---Courant Institute of Mathematical Sciences, 251 Mercer Street, New York, NY 10012 \\
              \email{davisad@alum.mit.edu}
           \and
           Brendan A. West \at
              Cold Regions Research and Engineering Laboratory, 72 Lyme Rd, Hanover, NH 03755
           \and
           Nathanael J. Frisch \at
              Cold Regions Research and Engineering Laboratory, 72 Lyme Rd, Hanover, NH 03755
        \and
           Devin T. O'Connor \at
              Cold Regions Research and Engineering Laboratory, 72 Lyme Rd, Hanover, NH 03755
         \and
           Matthew D. Parno \at
              Cold Regions Research and Engineering Laboratory, 72 Lyme Rd, Hanover, NH 03755
}


\date{Received: date / Accepted: date}

\maketitle

\begin{abstract}
ParticLS (\emph{Partic}le \emph{L}evel \emph{S}ets) is a software library that implements the discrete element method (DEM) and meshfree methods. ParticLS tracks the interaction between individual particles whose geometries are defined by level sets capable of capturing complex shapes. These particles either represent rigid bodies or material points within a continuum. Particle-particle interactions using various contact laws numerically approximate solutions to energy and mass conservation equations, simulating rigid body dynamics or deformation/fracture. By leveraging multiple contact laws, ParticLS can simulate interacting bodies that deform, fracture, and are composed of many particles. In the continuum setting, we numerically solve the peridynamic equations---integro-differential equations capable of modeling objects with discontinuous displacement fields and complex fracture dynamics. We show that the discretized peridynamic equations can be solved using the same software infrastructure that implements the DEM. Therefore, we design a unique software library where users can easily add particles with arbitrary geometries and new contact laws that model either rigid-body interaction or peridynamic constitutive relationships. We demonstrate ParticLS' versatility on test problems meant to showcase features applicable to a broad selection of fields such as tectonics, granular media, multiscale simulations, glacier calving, and sea ice.


\end{abstract}

\section{Introduction}
\label{sec:intro}

\sloppy Small-scale processes, such as the interaction of sand grains or fracture dynamics, impact the large-scale behavior of many materials. We devise a new software library that numerically simulates such phenomena using meshless methods or the discrete element method (DEM). These techniques solve equations derived from conservation of mass and momentum and constitutive models relating material deformation and stress. 
Meshless methods, like peridynamics \cite{Sillingetal2007}, smoothed-particle hydrodynamics \cite{Monaghan1992}, or reproducing kernel particle methods \cite{Bessaetal2014}, use partial differential equations or integro-differential equations to describe the material behavior. Alternatively, the DEM explicitly models small-scale behavior through the motion and interaction of discrete particles (e.g., sand grains) \cite{Andradeetal2012GranularElement,CundallStrack1979,GuoCurtis2015,Hopkins1991,Hopkins2004}.
Although DEM and meshless methods have fundamentally different material representations, we use abstract computational structures to implement both methods in a single software library, called ParticLS (\emph{Partic}le \emph{L}evel \emph{S}ets).
We will refer to both the DEM and peridynamics as ``meshfree'' techniques to highlight their similar computational structure. ParticLS is uniquely designed as an object-oriented software library that leverages software concepts such as abstraction and inheritance to organize and structure the code. Therefore, ParticLS provides core functionality that users can use to simulate complex geophysical systems such as sea ice or soils. 

The macroscale properties of granular media require simulating computationally expensive interactions between irregularly shaped particles \cite{Choetal2006,GuoMorgan2004}. Methods in the literature for representing complex shapes include clumping spherical particles \cite{Garciaetal2009}, 3-dimensional polyhedra \cite{Lathametal2001,LathamMunjiza2004}, tablet shaped particles \cite{Songetal2006}, or axisymmetrical particles \cite{Favieretal1999}.  
However, computing particle-particle contact information (e.g., the location where two particles collide) becomes computationally expensive given many particles or complex geometries \cite{Johnsonetal2004,Songetal2006}. Alternatively, irregular particle boundaries can be represented using the zero level set of a signed distance function (SDF) \cite{Kawamotoetal2016,Kawamotoetal2018}.

ParticLS uses this level set approach, as formulated by \cite{Kawamotoetal2016,Kawamotoetal2018}, which allows us to formulate a novel contact detection scheme as a constrained optimization problem. The contact point between a home particle and its neighbor is a point on the home particle's boundary that minimizes the neighbor particle's SDF. This generalized approach allows ParticLS to represent many unique particle geometries within one simulation without having to specify different contact models for every potential particle geometry combination. Users only need to define an SDF to add a new particle geometry that works with this contact detection algorithm.


ParticLS can also model deformable bodies using peridynamics, which can simulate fracturing and discontinuous materials \cite{Fosterestal2010,MadenciOterkus2016}. Peridynamics replaces the divergence operator on the Cauchy stress with an integral operator; these equations are valid even along crack tips and discontinuities where the spatial derivatives used to compute stress in traditional continuum formulations generate singularities \cite{Askarietal2008,Rogula1982,Silling2000,Sillingetal2007,SillingLehoucq2010,SillingLehoucq2008}. Peridynamic implementations share many features with the DEM and we leverage this similarity to provide a software framework for both approaches. 

ParticLS's object-oriented framework mimics mathematical formulations, allowing users to easily add functionality specific to their application. In this paper, we highlight four major contributions of ParticLS:
\begin{enumerate}
    \item We address computational limitations when computing contact between particles that have complex geometries by deriving an \textit{optimization-based contact detection algorithm} between neighboring particles.
    \item Generic meshfree algorithms (e.g., contact detection) are easily adapted to specific applications using abstract base classes that provide default implementations for important functionality. Users can easily add to the library by inheriting the functionality of these classes and prescribing their own definitions.
    \item Users can couple DEM and peridynamic methods with other time-dependent simulations or data. ParticLS uses state variables that are defined by an ordinary differential equation, such as temperature-dependent material properties or contact shear forces.  
    \item Deformable bodies can interact by implementing DEM and peridynamic models with the same code.
\end{enumerate}

Section \ref{sec:mathematical_formulation} describes the model formulation for both the DEM and peridynamics, specifically highlighting how the discretized peridynamic equations are analogous to a DEM. Section \ref{sec:contact-detection} describes the optimization-based contact detection algorithm. Section \ref{sec:object-oriented-implementation} describes ParticLS' object-oriented framework and Section \ref{sec:examples} shows selected examples to demonstrate ParticLS' unique capabilities.

\section{DEM and peridynamic background} \label{sec:mathematical_formulation}

\subsection{Discrete element method (DEM)}

The equations for the DEM are derived from force balance and conservation of angular momentum \cite{CundallStrack1979}. ParticLS's DEM implementation evolves the global position and orientation of a specific particle relative to a reference position. We use lowercase symbols and letters to denote quantities in the global frame and uppercase symbols to represent quantities in the local reference frame. Let $X$ denote the $d \in \{2, 3\}$ dimensional coordinate in a reference frame such that the particle's center of mass is at $X=0$. In this frame, a particle's geometry is defined by a SDF $\varphi: \mathbb{R}^{d} \mapsto \mathbb{R}$ \cite{Kawamotoetal2016,Kawamotoetal2018} such that 
\begin{equation}
    \varphi(X) \begin{cases}
    & < 0 \mbox{ if $X$ is inside the particle} \\
    & = 0 \mbox{ if $X$ is on the particle's boundary} \\
    & > 0 \mbox{ if $X$ is outside the particle.}
    \end{cases}
    \label{eq:signed-distance-function}
\end{equation}
Furthermore, the magnitude $|\varphi(X)|$ defines the minimum distance between $X$ and a point on the boundary. For the purpose of this paper, we assume that the particles are convex.
Let $\bar{y}_i(t)$ be the center of mass of particle $i \in \{1,2,\hdots,n\}$ in a global coordinate frame at time $t$ and $u_i(t)$ be the displacement at time $t$ such that $u_i(0)=0$ and $\bar{y}_i(t) = \bar{x}_i+u_i(t)$, where $\bar{x}_i$ is the center of mass at $t=0$. We employ quaternions to represent the orientation of the particles, which can be defined in terms of a unit orientation vector $\chi_i(t)$ ($\|\chi_i(t)\|_2=1$) and angle $\theta_i(t)$ 
\begin{equation}
    \begin{split}
        \mathpzc{q}_i(t) = & \cos{\left(\frac{\theta_i(t)}{2}\right)} + \\ & \left(\chi_i^{(0)}(t) \hat{\imath}
        + \chi_i^{(1)}(t) \hat{\jmath}+ \chi_i^{(2)}(t) \hat{k}\right) \sin{\left(\frac{\theta_i(t)}{2}\right)},
    \end{split}
\end{equation}
where $\chi_i^{(k)}(t)$ is the $k^{\text{th}}$ component of $\chi_i(t)$, such that rotating the particle about $\chi_i(t)$ by $\theta_i(t)$ radians aligns the global and reference coordinate frame. The transformation 
\begin{equation}
    X(x, \bar{y}_i(t), \mathpzc{q}_i(t)) = \mathpzc{q}_i(t) \circ (x-\bar{y}_i(t))
    \label{eq:global-to-reference-transformation}
\end{equation}
maps the global coordinate $x$ into the $i^{\text{th}}$ particle's local reference frame. The $\circ$ operator is the action of rotating the vector $x$ by the quaternion $\mathpzc{q}$.

The particles are translated and rotated in the global frame according to Newton's second law and conservation of angular momentum. Given the translational velocity $v_i(t) = \dot{u}_i(t)$ where the dot denotes time derivatives, 
\begin{subequations}
\begin{eqnarray}
    \dot{u}_i(t) &=& v_i(t) \\
    m_i \ddot{u}_i(t) = m_i \dot{v}_i(t) &=& \sum_{j=1}^{n} f_{ij}(t) + b_i(t),
\end{eqnarray}
\label{eq:dem-force-balance}%
\end{subequations}
where $m_i$ is the mass of particle $i$, $f_{ij}(t)$ is the force acting on particle $i$ from particle $j$, and $b_i(t)$ are body forces acting on particle $i$ (e.g., gravity). For particle $i$, let the vector $\omega_i(t)$ be the angular velocity, which defines the pure quaternion $\mathpzc{w}_i(t) = (0, \omega_i(t))$. The angular velocity describes the change in orientation 
\begin{subequations}
\begin{equation}
    \dot{\mathpzc{q}}_i(t) = \frac{1}{2} \mathpzc{w}_i(t) * \mathpzc{q}_i(t),
    \label{eq:dem-3d-quaternion-evolution}
\end{equation}
and evolves via conservation of angular momentum
\begin{equation}
    I_i(t) \dot{\omega}_i(t) = \sum_{j=1}^{n} \tau_{ij}(t) - \dot{I}_i(t) \omega_i(t),
\end{equation}
\label{eq:dem-rotational-equations}%
\end{subequations}
where $I(t)$ and $\dot{I}_i(t)$ is the moment of inertia for particle $i$ in the global frame and its time derivative and $\tau_{ij}(t)$ is the torque acting on particle $i$ from particle $j$. In two dimensions, the rotational equations \eqref{eq:dem-rotational-equations} significantly simplify because the particle is restricted to the $x$-$y$ plane, and therefore, we only need the orientation angle $\theta_i(t)$. The angular velocity is now also a scalar quantity such that \eqref{eq:dem-3d-quaternion-evolution} is replaced by $\dot{\theta}_i(t) = \omega_i(t)$.

\subsection{Peridynamic formulation}

\begin{figure}[ht]
 \begin{center}
 \includegraphics[scale=1.0]{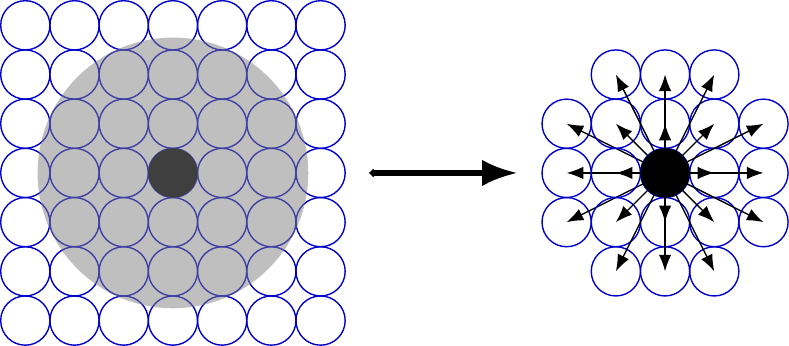}
 \end{center}
 \caption{Illustration of the non-local forces acting on the middle particle in a peridynamic setting. Each circle represents the material point $\mathcal{B}_r(x^{\prime})$. The gray region corresponds to the horizon $\mathcal{B}_{\delta}(x)$ of the middle particle. Each arrow in the right image corresponds to the pair-wise force per volume squared $\eta_{i}(x,x_{i},t)$ contributing to the force on the middle particle after discretization.}
 \label{fig:peridynamics-diagram}
\end{figure}

In the peridynamic framework, non-local forces acting across finite distances allow discontinuities and cracks to form within continuous materials \cite{Silling2000,Sillingetal2007}. Let $\mathcal{B}_\delta(x)$ be a ball with radius $\delta$ centered at $x$. Forces are modeled as bonds between a point $x$ and points in its horizon $\mathcal{B}_\delta(x)$, as illustrated in Figure \ref{fig:peridynamics-diagram}.

Given a material with density $\rho(x)$, the peridynamic equation is
\begin{equation}
    \rho(x) \ddot{u}(x, t) = \int_{\mathcal{B}_\delta(x)} \eta(x, x^{\prime}, t) \, dx^{\prime} + \beta(x, t),
    \label{eq:peridynamic-equation}
\end{equation}
where $u(x,t)$ is the unknown displacement, $\eta(x, x^{\prime}, t)$ is the pairwise force per volume squared, and $\beta(x,t)$ is the body force per volume. Mechanical restrictions require that
\begin{subequations}
\begin{eqnarray}
    && \eta(x, x^{\prime}, t) = -\eta(x^{\prime}, x, t), \\
    && \eta(x, x^{\prime}, t)=0 \mbox{ when }\|x-x^{\prime}\|_2>\delta,
\end{eqnarray}
\label{eq:peridynamic-force-density-assumption}
\end{subequations}
and
\begin{equation}
    \int_{\Omega} y(x, t) \times \int_{\mathcal{B}_{\delta}(x)} \eta(x, x^{\prime}, t) \, dx^{\prime} \, dx = 0,
    \label{eq:peridynamic-angular-momentum-conservation-assumption}
\end{equation}
where $y(x,t) = x + u(x,t)$ is the deformation \cite{Silling2000,Sillingetal2007}. These restrictions are sufficient conditions to conserve linear and angular momentum. To discretize, we first integrate over a ball centered at material point $x_i$ with radius $r \ll \delta$
\begin{equation}
    \begin{split}
        \int_{\mathcal{B}_r(x_i)} \rho(x) & \ddot{u}(x, t) \, dx = \\ & \int_{\mathcal{B}_r(x_i)} \int_{\mathcal{B}_\delta(x)} \eta(x, x^{\prime}, t) \, dx^{\prime} + \beta(x, t) \, dx.
    \end{split}
\end{equation}
Let the density $\rho_i = \rho(x_i)$ and assume that it is approximately constant in $\mathcal{B}_r(x_i)$. We also assume that the discrete force per volume squared $\eta_{ij}(t) = \eta(x_i, x_j, t)$ is approximately constant with respect to the position $x$ (or $x^{\prime}$) in the ball $\mathcal{B}_r(x)$ (or $\mathcal{B}_r(x^{\prime})$) and that $\eta_{ij}(t)=0$ if $\|x_i-x_j\|>\delta$. Similarly, we assume that the acceleration $\ddot{u}_i(t) = \ddot{u}(x_i, t)$ and body force density $\beta_i(t) = \beta(x_i, t)$ are constant in $\mathcal{B}_r(x_i)$. The discrete peridynamic equation is, therefore,
\begin{equation}
    \rho_i V_r \ddot{u}_i(t) \approx \sum_{j=1}^{n} \eta_{ij}(t) V_r^2 + \beta_i(t) V_r,
    \label{eq:discrete-peridynamic-equation}
\end{equation}
where $V_r$ is the volume of a $r$-radius ball. Defining $m_i = \rho_i V_r$, $f_{ij}(t) = \eta_{ij}(t) V_r^2$, and $b_i(t) = \beta_i(t) V_r$ derives the peridynamic analog to \eqref{eq:dem-force-balance}. The restriction on $\eta(x, x^{\prime}, t)$ requiring \eqref{eq:peridynamic-angular-momentum-conservation-assumption} ensures that the system conserves angular momentum and, therefore, we do not need to explicitly evolve the orientation of each particle \cite{Sillingetal2007}.

\subsubsection{State-based peridynamics}

ParticLS implements state-based peridynamics, which is an extension of the formulation above but a detailed description is beyond the scope of this paper---see \cite{Sillingetal2007} for details. Briefly, state-based peridynamic theory defines the pairwise force per squared volume $\eta(x, x^{\prime}, t)$ via a force vector state $F[x,x^{\prime},t]\langle x^{\prime}-x \rangle$, which is a potentially nonlinear and discontinuous operator $F[x,x^{\prime},t]$ that acts on the bond $\xi = x^{\prime}-x$ \cite{Sillingetal2007}. State-based peridynamics is the special case of \eqref{eq:peridynamic-equation} with
\begin{equation}
    \eta(x, x^{\prime}, t) = F[x, x^{\prime}, t]\langle x^{\prime}-x \rangle - F[x^{\prime}, x, t]\langle x-x^{\prime} \rangle.
    \label{eq:peridynamic-force-density}
\end{equation}
We assume forces only act across bonds, therefore, define
\begin{equation} 
    Y[x, x^{\prime}, t]\langle x^{\prime}-x \rangle = \frac{y(x^{\prime}, t) - y(x, t)}{\|y(x^{\prime}, t) - y(x, t)\|_2}
\end{equation}
to be the unit vector state that points across the deformed bond $y(x, t) - y(x^{\prime}, t) = (x^{\prime} + u(x^{\prime}, t)) - (x + u(x, t))$. The force state of ordinary materials is, therefore, written in terms of a scalar state $f(x, x^{\prime}, t)$
\begin{equation}
    F[x, x^{\prime}, t]\langle x^{\prime}-x \rangle  = f(x, x^{\prime}, t) Y[x,t]\langle x^{\prime}-x \rangle.
    \label{eq:ordinary-force-state}
\end{equation}
Given $f(x, x^{\prime}, t)$, \eqref{eq:ordinary-force-state} defines the force state that is substituted into \eqref{eq:peridynamic-force-density}.

\section{Optimization-based contact detection}
\label{sec:contact-detection}

As introduced in Section \ref{sec:intro}, ParticLS implements a novel contact detection routine that is formulated as an optimization problem based on the particles' SDFs. The LS-DEM, developed by \cite{Kawamotoetal2016}, stores discrete points along a particle's boundary (zero level set surface). Contact is detected by looping over these boundary points and determining if a neighbor's SDF is negative for each boundary point. This approach has two potential drawbacks: (i) it is computationally expensive to iterate through each set of boundary points for simulations with many particles and (ii) the contact force between the two particles can be sensitive to the number of points that are placed on the particle boundary \cite{li2019capturing}. Instead, ParticLS formulates an optimization problem that finds the point on the ``home'' particle that penetrates deepest into its neighbor. This contact routine is defined abstractly in terms of the SDF so that new particle geometries are easily added by defining the corresponding SDF.

Consider two particles defined by their centers of mass ($\bar{x}$ and $\bar{x}^{\prime}$), orientations ($\mathpzc{q}$ and $\mathpzc{q}^{\prime}$), and SDFs ($\varphi(X(x, \bar{x}, \mathpzc{q})$ and $\varphi^{\prime}(X(x, \bar{x}^{\prime}, \mathpzc{q}^{\prime})$); denote these particles $p = (\bar{x}, \mathpzc{q}, \varphi)$ and $p^{\prime} = (\bar{x}^{\prime}, \mathpzc{q}^{\prime}, \varphi^{\prime})$. We refer to a `point of contact' as the location on one particle's boundary that is closest to the neighboring particle, or the deepest point of penetration if the particles are penetrating each other. A potential point of contact is located at a point $x_1$ on the boundary of particle $p$ that is closest to particle $p^{\prime}$.\footnote{Although we have assumed that particles are convex, this method is compatible for non-convex particles with minimal alterations. In the non-convex case, multiple contact points are possible and each potential contact point corresponds to a local minimum of Equation \eqref{eq:bond-optimization}. We can identify these local solutions by repeatedly solving the optimization problem with different initial guesses. The details of the non-convex case are beyond the scope of this paper and we, therefore, leave further discussion to future work.} There is a corresponding point $x_2$ on the boundary of particle $p^{\prime}$ that minimizes the distance to particle $p$.  Mathematically, these contact points solve the optimization problem
\begin{equation}
\begin{aligned}
& \underset{x_1,x_2 \in \mathbb{R}^{d}}{\text{minimize}}
& & \varphi(X(x_2, \bar{x}, \mathpzc{q}))+\varphi^{\prime}(X(x_1, \bar{x}^{\prime}, \mathpzc{q}^{\prime})) \\
& \text{subject to}
& & \varphi(X(x_1, \bar{x}, \mathpzc{q})) = 0\\
& &  &\varphi^{\prime}(X(x_2,  \bar{x}^{\prime}, \mathpzc{q}^{\prime})) = 0.
\end{aligned}
\label{eq:bond-optimization}
\end{equation}
{\color{EDIT_COLOR}{In principle, we can use any standard optimization algorithm to solve this problem. In some cases, such as sphere-sphere contact, this problem has a known analytical solution. In others, we can leverage the specific form of the signed distance function to devise a specialized algorithm; for example, the Gauss-Seidel can efficiently solve this problem for polygon-polygon contact. More generally, when we do not have an analytical solution or specialized algorithm we solve this optimization problem using the alternating direction method of multipliers (ADMM)---also called Douglas-Rachford splitting \cite{Benningetal2015,Goldsteinetal2014,ParikhBoyd2014}}. Importantly, ADMM does not require derivative information of the SDF, instead relying on the proximal operator, which allows particle geometry to have sharp corners. Since the SDF is continuous by definition, this algorithm does not impose any restrictions on the particle geometry. A detailed description of these algorithms is beyond the scope of this paper.}

{\color{EDIT_COLOR}{
Previous ``brute-force'' contact detection algorithms, which stores $n$ points along the zero level set, are often significantly more computationally expensive than solving the optimization problem posed in Equation \eqref{eq:bond-optimization}. Specifically, in the brute force algorithm, the ``home'' particle needs to evaluate its neighbor's SDF at all $n$ points on its boundary. In comparison, ADMM typically converges very quickly. We also use the solution from the previous timestep as an initial condition, which often leads to convergence in 1-2 iterations. Additionally, ADMM finds the deepest penetration point up to a user-specified tolerance, rather than approximating this point using a pre-defined set along the particle boundary. Finally, storing the required $n$ boundary points for every particle in the simulation can require a huge amount of computer memory for simulations with many thousands of particles. 
}}

For convex particles, the points $x_1$ and $x_2$ are unique \cite{ParikhBoyd2014} and represent \textit{potential} contact locations. The particles are penetrating if both distance functions are negative.\footnote{Note that the SDFs will always have the same sign.} When this occurs, it is useful to look at the bond $\xi$, which is the vector pointing from the first particle to the second, as shown in Figure \ref{fig:particle-geometry}. Mathematically, the bond is defined by
\begin{subequations}
\begin{equation}
    \xi = x_2-x_1
    \label{eq:bonds}
\end{equation}
and the bond length is $\|\xi\|_2$. When the particles are penetrating, the bond length is the penetration distance. Given the translational velocities $v(t)$ and $v^{\prime}(t)$, the relative velocity is 
\begin{equation}
    v_{r}(t) = (v(t)+\omega \times (x-x_1))-(v^{\prime}(t)+\omega^{\prime} \times (x^{\prime}-x_2))
\end{equation} 
and the tangential velocity, as defined by \cite{Kawamotoetal2016}, is
\begin{equation}
    v_a(t) = v_r(t) - (v_r(t) \cdot \xi) \xi / \|\xi\|_2.
    \label{eq:tangential-velocity}
\end{equation}
\label{eq:contact-summary}%
\end{subequations}
The torque and force between the two particles depends on the contact summary as defined by Equation \eqref{eq:contact-summary}.
{\color{EDIT_COLOR}{For non-convex particles, we divide their geometry into convex sub-particles and then find the potential contact points between each sub-particle and the neighbor. The resultant set of points describes potential contact points for the non-convex particles, and allows for multiple contact points, which we expect to happen in certain non-convex contact scenarios.}}

\newcommand{\normal}[3]{
    \draw[-latex, thick] (#2) -- ($(#1)!(#2)!(#3)!1.5!(#2)$);
}
\tikzset{
  arrowblack/.style={{Circle[black,length=2pt]}-Latex, color=black}
}

\begin{figure}[ht!]
  \begin{center}
 \begin{tikzpicture}
    \begin{scope}[rotate=-45]
    \path[font={\tiny}]
        (1.5, 0)   coordinate (p1A)
        (0, -1)   coordinate (p1B) 
        (0, 1.5)   coordinate (p1C)
        (1.5, 1.5)   coordinate (p1D)
    ;
    \draw[red!80!black, thick] plot [smooth cycle, tension=0.8] coordinates {(p1A) (p1B) (p1C) (p1D)};
    
     \path[font={\tiny}]
        (3.5, 2)   coordinate (p2A)
        (2, 2)   coordinate (p2B)
        (2, 3.5)   coordinate (p2C)
        (3.5, 3)   coordinate (p2D) 
    ;
    \draw[blue!80!black, thick] plot [smooth cycle, tension=0.8] coordinates {(p2A) (p2B) (p2C) (p2D)};
    
    \draw[arrowblack] (p1D) -- (p2B);
    \end{scope}
    
    \node[rectangle split, rectangle split parts=2, align=left] at (1.9,-0.2) (p1) {$x_1$};
    \node[rectangle split, rectangle split parts=2, align=left] at (3.1,-0.2) (p1) {$x_2$};
    
    \begin{scope}[rotate=-45,shift={(2.75,-2.75)}]
    \path[font={\tiny}]
        (1.5, 0)   coordinate (p1A)
        (0, -1)   coordinate (p1B) 
        (0, 1.5)   coordinate (p1C)
        (1.5, 1.5)   coordinate (p1D)
    ;
    \draw[red!80!black, thick] plot [smooth cycle, tension=0.8] coordinates {(p1A) (p1B) (p1C) (p1D)};
    
     \path[font={\tiny}]
        (2.5, 1)   coordinate (p2A)
        (1, 1)   coordinate (p2B)
        (1, 2.5)   coordinate (p2C)
        (2.5, 2)   coordinate (p2D) 
    ;
    \draw[blue!80!black, thick] plot [smooth cycle, tension=0.8] coordinates {(p2A) (p2B) (p2C) (p2D)};
    
    \draw[arrowblack] (p1D) -- (p2B);
    \end{scope}
    
    \node[rectangle split, rectangle split parts=2, align=left] at (2.4,-4.1) (p1) {$x_1$};
    \node[rectangle split, rectangle split parts=2, align=left] at (1.2,-4.1) (p1) {$x_2$};

    \node[rectangle split, rectangle split parts=2, align=left] at (0.1,-2.25) (p1) {\color{red!80!black}{Particle $p=(\bar{x}, \mathpzc{q}, \varphi)$}};
    \node[rectangle split, rectangle split parts=2, align=left] at (4.5,-2.25) (p2) {\color{blue!80!black}{Particle $p^{\prime}=(\bar{x}^{\prime}, \mathpzc{q}^{\prime}, \varphi^{\prime})$}};

 \end{tikzpicture}
 \end{center}
 \caption{The red and blue curves correspond to $\varphi(X(x_1, \bar{x}, \mathpzc{q}))=0$ and $\varphi^{\prime}(X(x_1, \bar{x}^{\prime}, \mathpzc{q}^{\prime}))=0$ level sets of two particles $p=(\bar{x}, \mathpzc{q}, \varphi)$ and $p^{\prime}=(\bar{x}^{\prime}, \mathpzc{q}^{\prime}, \varphi^{\prime})$. The arrows represent the bond $\xi=x_1-x_2$.}
 \label{fig:particle-geometry}
\end{figure}
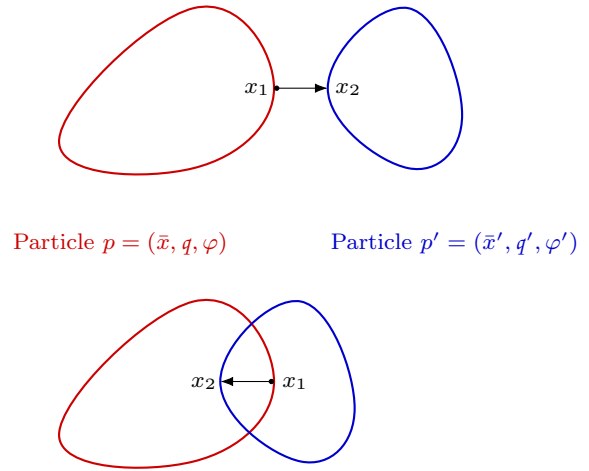

\subsection{Example: elastic contact} \label{sec:elastic-contact}

Elastic contact highlights an example where the force and torque are parameterized by the contact summary \eqref{eq:contact-summary} \cite{CundallStrack1979}. Given two particles $p(t)=(\bar{x}(t), \bar{\mathpzc{q}}(t), \varphi)$ and $p^{\prime}(t)=(\bar{x}^{\prime}(t), \bar{\mathpzc{q}}^{\prime}(t), \varphi^{\prime})$, the normal force is 
\begin{equation}
    F_{n,e}(t) = -\kappa_n \llbracket \varphi^{\prime}(X(x_1, \bar{x}^{\prime}(t), \mathpzc{q}^{\prime}(t))) < 0 \rrbracket \xi,
    \label{eq:elastic-normal}
\end{equation}
where $\llbracket \cdot \rrbracket$ is the indicator function. The shear force evolves according to
\begin{equation}
    \dot{F}_{s,e}(t) = - \kappa_s \llbracket \varphi^{\prime}(X(x_1, \bar{x}^{\prime}(t), \mathpzc{q}^{\prime}(t))) < 0 \rrbracket
     v_a(t)
    \label{eq:elastic-shear-evolution}
\end{equation}
with $F_{s,e}(0) = 0$. If the particles are not in contact then we reset $F_{s,e}=0$---this is important if the particles ever deflect away from each other then come back into contact. A Coulomb friction criterion is applied to the shear force according to 
\begin{equation}
    F^{c}_{s,e}(t) = \left\{
                    \begin{array}{ll}
                        F_{s,e}(t) & \text{if } \|F_{s,e}(t)\| \leq \mu \|F_{n,e}(t)\|\\
                        \mu F_{n,e}(t) & \text{otherwise}
                    \end{array}
                \right.
    \label{eq:coulomb-friction}
\end{equation}
where $\mu$ is the friction coefficient. The total force acting on particle $p$ due to particle $p^{\prime}$ is therefore $F(t) = F_{n,e}(t)+F^{c}_{s,e}(t)$ and the torque is $\tau(t) = ((x_1+x_2)/2-\bar{x}) \times F(t)$. In this example, we define the moment arm $(x_1+x_2)/2$ to be halfway between the points that solve \eqref{eq:bond-optimization}. In two dimensions, this simplifies by restricting all translation and rotation to the $x$-$y$ plane. 

\subsection{Example: viscoelastic contact} \label{sec:viscoelastic-contact}

We also highlight linear viscoelastic contact, which can be conceptualized as a linear spring-dashpot system \cite{Shafer1996}. This model is similar to the elastic contact model but with additional viscous damping components that remove energy from particle interactions. The normal force due to damping is 
\begin{equation}
    F_{n,v}(t) = -\gamma_n v_r(t),
    \label{eq:viscelastic-normal}
\end{equation}
where $\gamma_n$ is the normal viscous damping coefficient. Therefore the total normal force is $F_{n}(t) = F_{n,e}(t)+F_{n,v}(t)$. The same evolution equation \eqref{eq:elastic-shear-evolution} models the shear force, however, viscoelastic contact adds an additional shear force component that accounts for damping:
\begin{equation}
    F_{s,v}(t) = -\gamma_s v_a(t),
    \label{eq:viscelastic-shear}
\end{equation}
where $\gamma_s$ is the tangential viscous damping coefficient. Therefore, the total normal force is $F_{s}(t) = F_{s,e}(t)+F_{s,v}(t)$. The total shear force applies the same Coulomb friction criterion in \eqref{eq:coulomb-friction}. The torque remains $\tau(t) = ((x_1+x_2)/2-\bar{x}) \times F(t)$. 

\section{ParticLS object-oriented implementation} \label{sec:object-oriented-implementation}

ParticLS defines abstract objects that dictate an interface to important functionality for meshfree methods. Through inheritance and abstraction users can easily customize base functionality to implement new particle geometries, body forces, contact forces, etc.. This makes ParticLS a useful tool for a wide variety of applications, e.g. soil mechanics, fracturing bodies, and/or sea ice. 
We refer to objects/data structures as ``classes" and denote them with the \texttt{typewritter} font. 

ParticLS evolves time-dependent quantities of interest by evolving a collection of abstract objects called \texttt{StateObject}s, which are stored in the \texttt{Simulation}. The simulation-level state is the combined state of its \texttt{StateObject}s. ParticLS compartmentalizes \texttt{Simulation} into (i) a collection of time-dependent \texttt{StateObject}s, (ii) numerical algorithms required to evolve the states (finding particle nearest neighbors and time-integration), and (iii) observers that extract quantities of interest at specified times. 

\subsection{Collections of time-dependent state objects}

Unlike other particle methods, this allows ParticLS to additionally evolve non-particle states---for example, time-dependent geometries. Each \texttt{StateObject} stores a time-dependent parameter, such as a rigid body's position and velocity or the shear force defined by \eqref{eq:elastic-shear-evolution} and \eqref{eq:viscelastic-shear}. Rather than explicitly integrating the shear force, many algorithms use first-order finite difference to approximate an explicit expression for $F_s$. However, if we use a higher-order method to integrate particle position, the time-discretization of the shear force must also be updated---otherwise we lose higher-order accuracy in time. Rather than tie expressions for time-dependent quantities to their discretization, we incorporate them into the time-dependent state of the \texttt{Simulation} object via the \texttt{StateObject} structure. 

Rigid particles (material points in the peridynamic setting) are implemented as \texttt{RigidBody}s, which inherit from \texttt{StateObject}. \texttt{RigidBody} stores the
\begin{enumerate}
    \item center of mass (a $d$ dimensional vector $x_i(t)$),
    \item reference particle geometry (a pointer to a \texttt{Geometry} object),
    \item translational velocity (a $d$ dimensional vector $v_i(t)$),
    \item orientation (an angle $\theta_i(t)$ in two dimensions and a quaternion $\mathpzc{q}_i(t)$ in three dimensions), and 
    \item angular velocity (a scalar in two dimensions and a three dimensional vector in three dimensions $\omega_i(t)$).
\end{enumerate}
The global coordinate $x$ is transformed into the particle's reference coordinate $X$ via  \eqref{eq:global-to-reference-transformation} using the center of mass and orientation. Given the reference coordinate $X$, the \texttt{RigidBody} evaluates its \texttt{Geometry}'s SDF \eqref{eq:signed-distance-function} to determine its shape. The \texttt{Geometry} object also computes the particle's mass and moment of inertia, given density $\rho(X)$. The \texttt{RigidBody} object also contains vectors of \texttt{BodyForce}s and \texttt{ContactForce}s, which represent forces acting on the particle. Given the net force, torque, translational velocity, and angular velocity, \texttt{RigidBody} computes the time-derivative of its state using \eqref{eq:dem-force-balance} and \eqref{eq:dem-rotational-equations}. The \texttt{Simulation} object storing the \texttt{RigidBody}s uses their time derivatives to evolve in time. \texttt{Simulation} also evolves other time-dependent \texttt{StateObject}s, such as shear forces stored in \texttt{ContactForces}. 

\subsection{Algorithms for evolving simulations}

\subsubsection{Finding nearest neighbors with $k$-d trees}

The \texttt{Searcher} object finds the nearest neighbors for each particle in the simulation. We define ``nearest neighbors'' with a distance metric between particles $d(p_i, p_j)$ and say $p_i$ and $p_j$ are neighbors if $d(p_i, p_j)<h$. For reference, the worst case algorithm is to first loop over every particle, and for each of these `outer' particles we again loop over all of the particles and decide if they are neighbors. The overall complexity of the algorithm is therefore $O(n^2)$. Instead, we organize the particles in a $k$-d tree to significantly reduce the complexity of finding the nearest neighbors. Assuming $d(p_i, p_j) = \|\bar{x}_i-\bar{x}_j\|_2$, a $k$-d tree stores each particle as a node in a binary tree \cite{Bentley1975}. ParticLS uses the open source software package \textit{nanoflann} to construct $k$-d trees \cite{nanoflann}. On average, the complexity of constructing $k$-d trees is $O(n \log{(n)})$ and finding the nearest neighbors is $O(\log{(n)})$ per particle. Therefore, the overall complexity of constructing the tree and finding each particle's nearest neighbors is $O(2n \log{(n)})$. We note that this complexity estimate assumes that particles are relatively evenly distributed throughout the domain or that the search finds the $k$ nearest neighbors (rather than the neighbors within radius $\Delta$). It is also important to note that the $k$-d tree approach is limited by the Euclidean distance metric between points. 

\subsubsection{Integrating simulation state in time} \label{sec:time-integration}

ParticLS implements second order Runge-Kutta for all \texttt{StateObject}s except \texttt{RigidBody}s, which use the second order velocity Verlet. Given timestep $\Delta t$ and any \texttt{StateObject} that is not a \texttt{RigidBody}, the time discretization is 
\begin{subequations}
\begin{eqnarray}
    s_{t+\frac{1}{2}} &=& s_{t} + \frac{\Delta t}{2} f(s_{t}, t \Delta t) \\
    s_{t+1} &=& s_{t+\frac{1}{2}} + \frac{\Delta t}{2} f(s_{t+\frac{1}{2}}, t \Delta t + \Delta t/2),
\end{eqnarray}
\label{eq:velocity-Verlet-two-step}
\end{subequations}
where $f(s_{t}, \tau)$ is the result of computing the \texttt{StateObject}'s right hand side at state $s_{t}$ and time $\tau$. The \texttt{RigidBody}'s right hand side computes the time derivatives of $v_t$ and $\omega_t$, $\dot v_t$ and $\dot\omega_t$, respectively. This defines the velocity and acceleration of the \texttt{RigidBody} and, therefore, 
\begin{equation}
    x_{t+1} = x_{t} + v_t \Delta t + \dot v_t \Delta t^2 / 2
\end{equation}
is a second-order discretization. Similarly, {\color{EDIT_COLOR}we use $\omega_t$ and $\dot \omega_t$ to define a second order scheme to update the orientation quaternion \cite{lim2014granular}
\begin{eqnarray}
    \mathpzc{q}_{t+1} = \frac{\Delta t }{2} (\mathpzc{w}_t + \frac{\Delta t }{2} \dot{\omega}_t ) * \mathpzc{q}_t,
\end{eqnarray}
where $\omega_t$ is the angular velocity and $\mathpzc{w}_t$ is the corresponding pure quaternion. 
}

\subsection{Observers and monitoring}

ParticLS provides functionality for tracking different quantities of interest throughout a simulation. This capability makes it easy for users to measure time-series of a specific particle's properties, or quantities related to the whole particle collection. The base class for this functionality is the \texttt{Observer} class. The \texttt{ParticleTracker}, for example, tracks the state variables of a \texttt{RigidBody}, which allows the simulation to extract the position, velocity, orientation, and/or angular velocity of a particular particle-of-interest. Similarly, the \texttt{CollectionTracker} computes quantities that depend on the entire collection. For example, in the DEM setting, we can use the \texttt{CollectionTracker} observer to measure the homogenized Cauchy stress tensor of the collection:
\begin{equation}
    \sigma = \frac{1}{V} \sum_{i=1}^{n} \sum_{j=1}^{n} d_{ij} \otimes F_{ij},
    \label{eqn:cauchy-stress}
\end{equation}
where $V$ is the volume of the smallest axis-aligned bounding box containing all of the particles, $n$ is the number of particles in the collection, $d_{ij} = \bar{x}_j-\bar{x}_i$ is the vector pointing from particle $i$'s center of mass to particle $j$'s center of mass, and $F_{ij}$ is the force vector acting between particles $i$ and $j$ \cite{GuoZhao2014}. The operator $\otimes$ denotes the outer product. We note that $F_{ij} = 0$ for most pairs and that \eqref{eqn:cauchy-stress} can be computed much more efficiently by replacing the second summation with the nearest neighbors of particle $i$. In practice, this is how we compute the Cauchy stress tensor. The \texttt{Observer} class and its children (e.g., \texttt{ParticleTracker} and \texttt{CollectionTracker}) lets ParticLS monitor observable quantities that can be directly compared to observations.

\section{Example applications} \label{sec:examples} 

\subsection{Non-spherical particle geometries}

\begin{figure}[ht!]
\centering
\begin{minipage}{.385\linewidth}
  \centering
  \includegraphics[width=\linewidth]{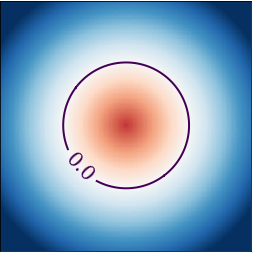}
\end{minipage}
\begin{minipage}{.385\linewidth}
  \centering
  \includegraphics[width=\linewidth]{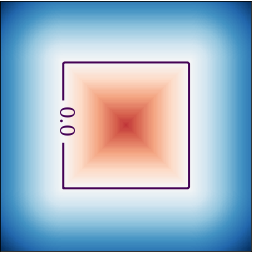}
\end{minipage}
\begin{minipage}{.385\linewidth}
  \centering
    \includegraphics[width=\linewidth]{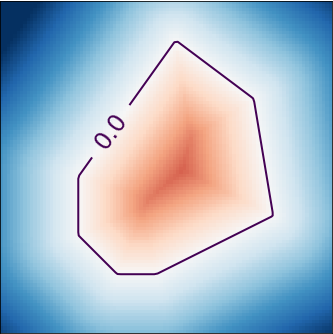}
\end{minipage}
\begin{minipage}{.385\linewidth}
  \centering
    \includegraphics[width=\linewidth]{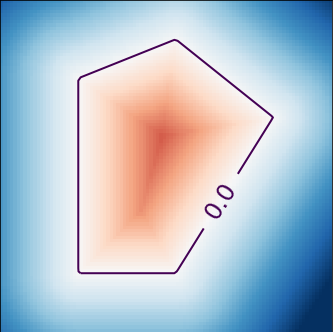}
\end{minipage}%
\caption{Top row: SDFs for the sphere and the box classes---\eqref{eq:SDF-sphere} and \eqref{eq:SDF-box}, respectively. Bottom row: SDFs for two different polygons.}

\label{fig:signed-distance-functions}
\end{figure}

Our first example highlights various particle geometries, which are implemented by creating children of the \texttt{Geometry} base class. Each new \texttt{Geometry} must compute its SDF in reference coordinates $X$ and, optionally, can implement computationally efficient functions to calculate its mass, moment of inertia, and other properties. Given this information, ParticLS can incorporate the new geometry into DEM or peridynamic simulations. We define three example shapes: (i) spheres, (ii) boxes, and (iii) polyhedra---see Figure \ref{fig:signed-distance-functions}. The SDF for a sphere with radius $R$ is 
 \begin{equation}
    \varphi(X) = \|X\|_2 - R.
    \label{eq:SDF-sphere}
\end{equation}
The SDF for a box is defined by the $d \in \{2, 3\}$ dimensional vector $Y$ such that $Y_i$ is the length of the box in the $i^{\mbox{th}}$ dimension. Considering the $d \in \{2, 3\}$ dimensional vector $D$ whose components are $D_{i} = \max{(\vert X_i \vert - Y_{i}/2), 0)}$; the SDF is
\begin{equation}
    \varphi(X) = \|D\|_2 - \min_{i \in \{1,\hdots,d\}}{(D_i)}.
    \label{eq:SDF-box}
\end{equation}

\begin{figure}[ht!]
\captionsetup[subfigure]{labelformat=empty}
\begin{center}
\begin{subfigure}[b]{0.325\linewidth}
 \includegraphics[width=\linewidth]{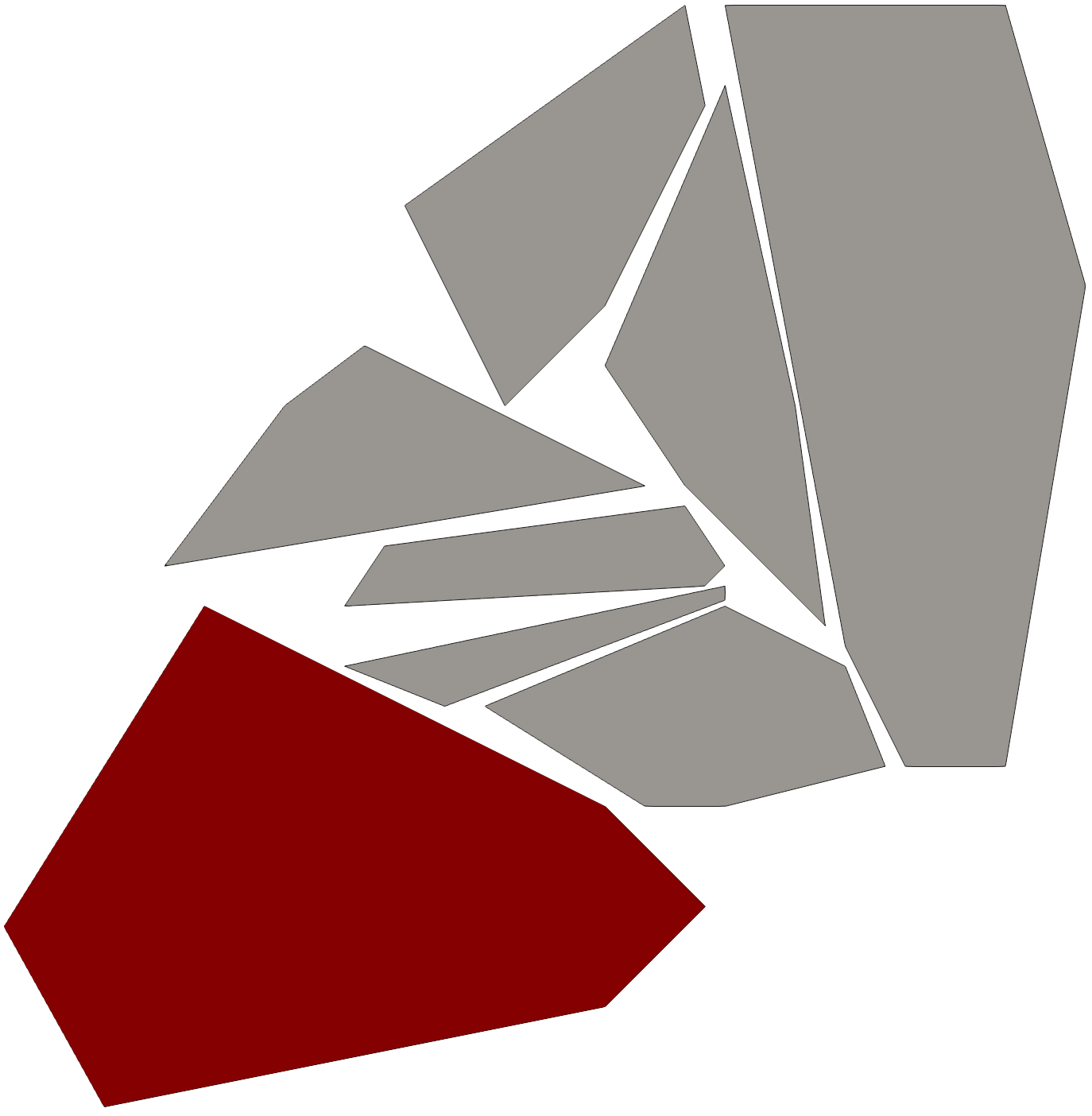}
 \caption{$t=0$}
\end{subfigure}
\begin{subfigure}[b]{0.325\linewidth}
 \includegraphics[width=\linewidth]{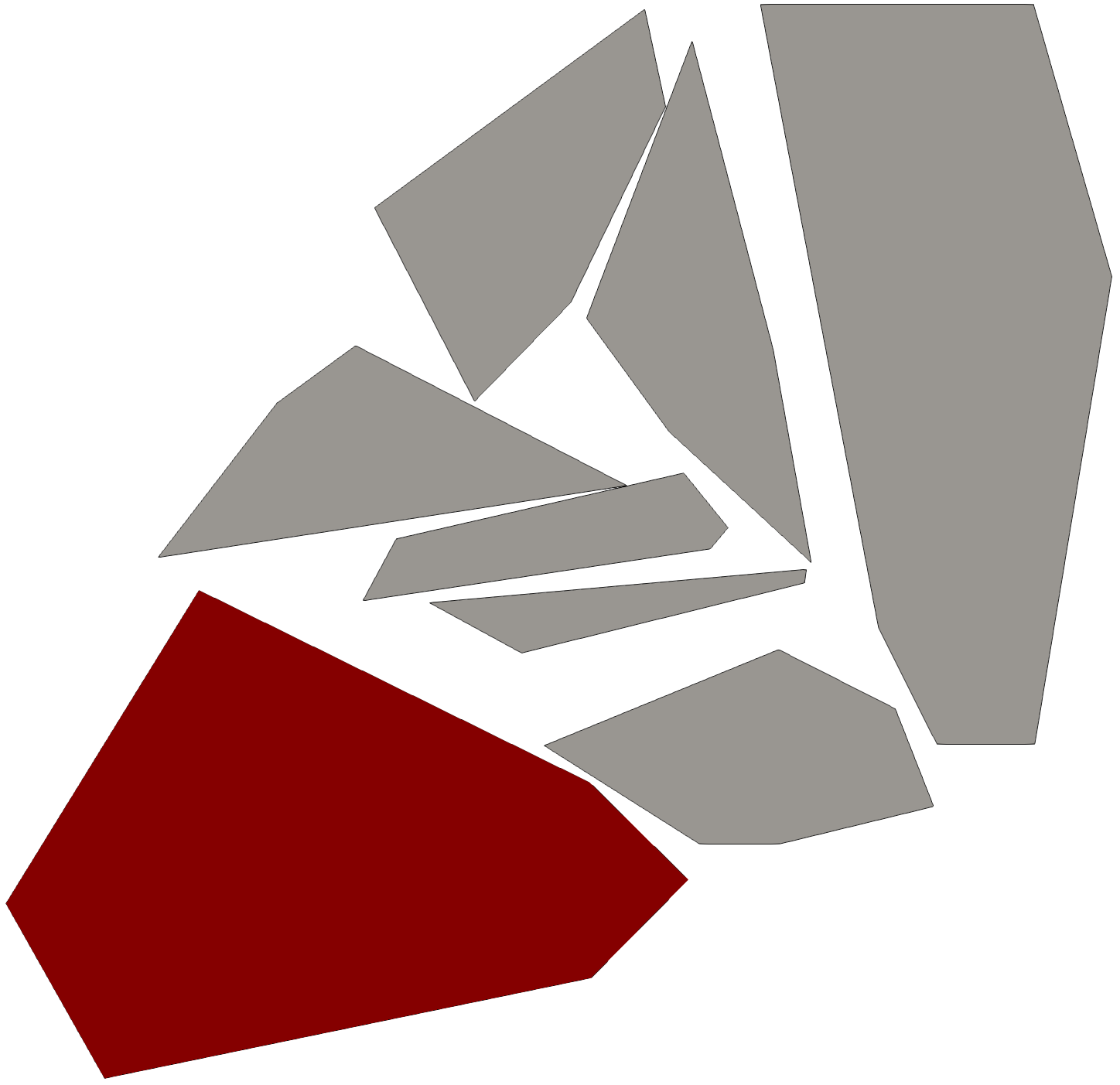}
 \caption{$t=0.5$}
\end{subfigure}
\begin{subfigure}[b]{0.325\linewidth}
 \includegraphics[width=\linewidth]{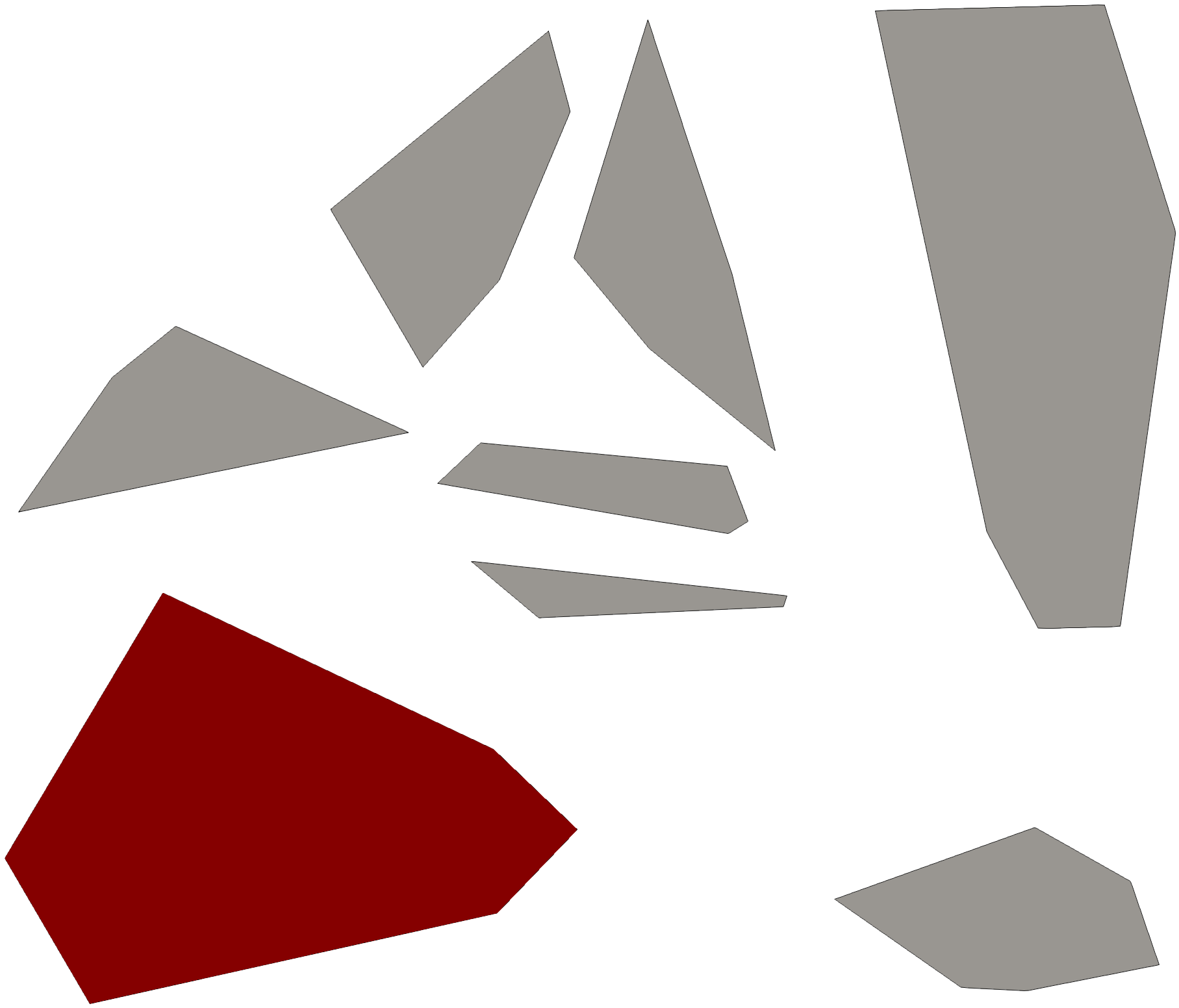}
 \caption{$t=1$}
\end{subfigure}
\end{center}
\caption{We define polygon geometries with density $\delta = 1$ that interact via the viscoelastic contact (section \ref{sec:viscoelastic-contact}) with $\kappa_n = 10^{5}$  $N/m$, $\kappa_s = 10^{4}$ $N/m$, $\mu = 1$, $\gamma_n = 1$ $Ns/m$, and $\gamma_s = \sqrt{0.1}$  $Ns/m$. We prescribe an initial velocity of $v_0 = (0.25, 0.24)$ $m/s$ to the red particle.}
\label{fig:polygon-contact}
\end{figure}

\begin{figure}[ht!]
\captionsetup[subfigure]{labelformat=empty}
\begin{center}
\begin{subfigure}[b]{0.325\linewidth}
 \includegraphics[width=\linewidth]{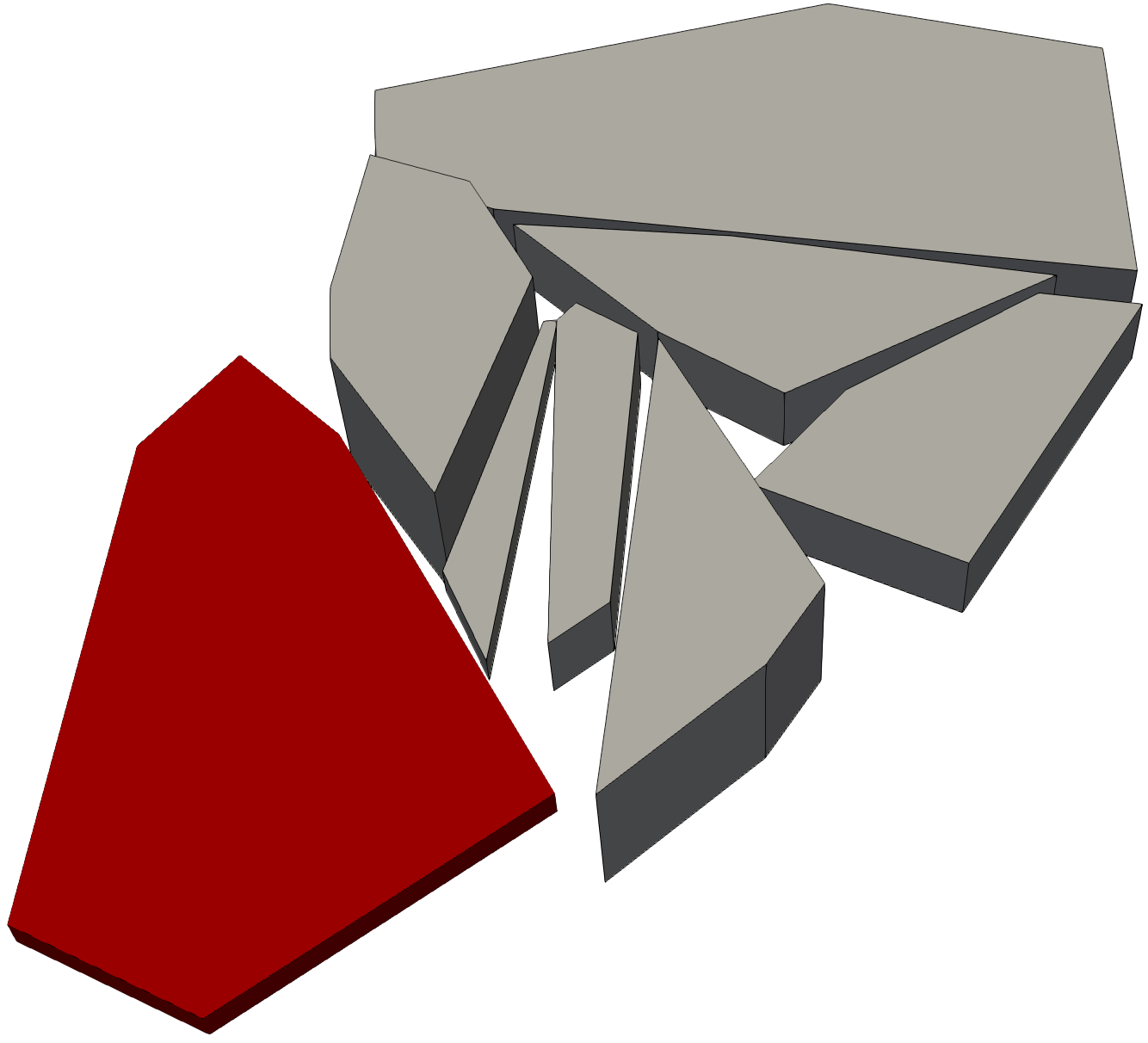}
 \caption{$t=0$}
\end{subfigure}
\begin{subfigure}[b]{0.325\linewidth}
 \includegraphics[width=\linewidth]{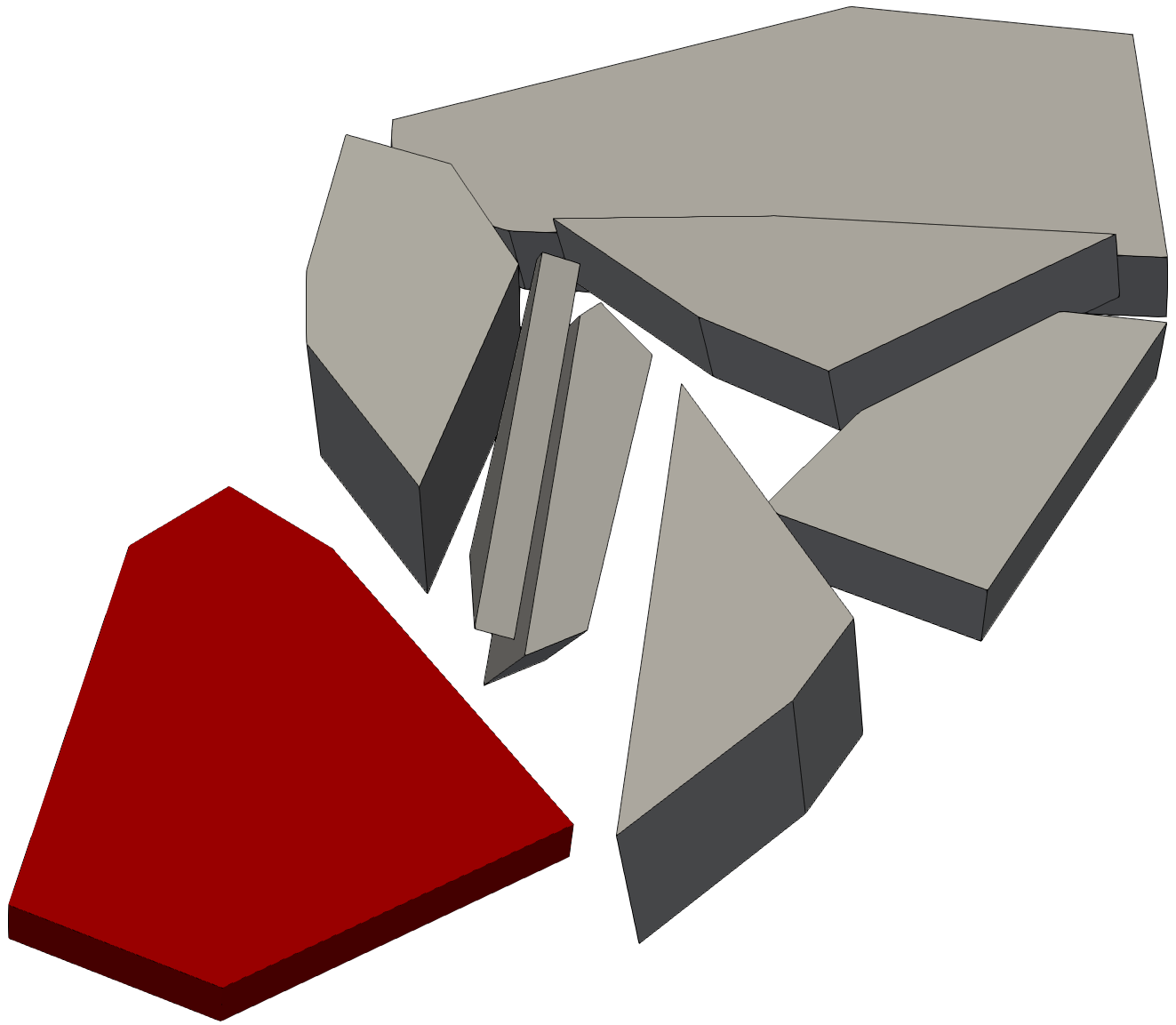}
 \caption{$t=0.5$}
\end{subfigure}
\begin{subfigure}[b]{0.325\linewidth}
 \includegraphics[width=\linewidth]{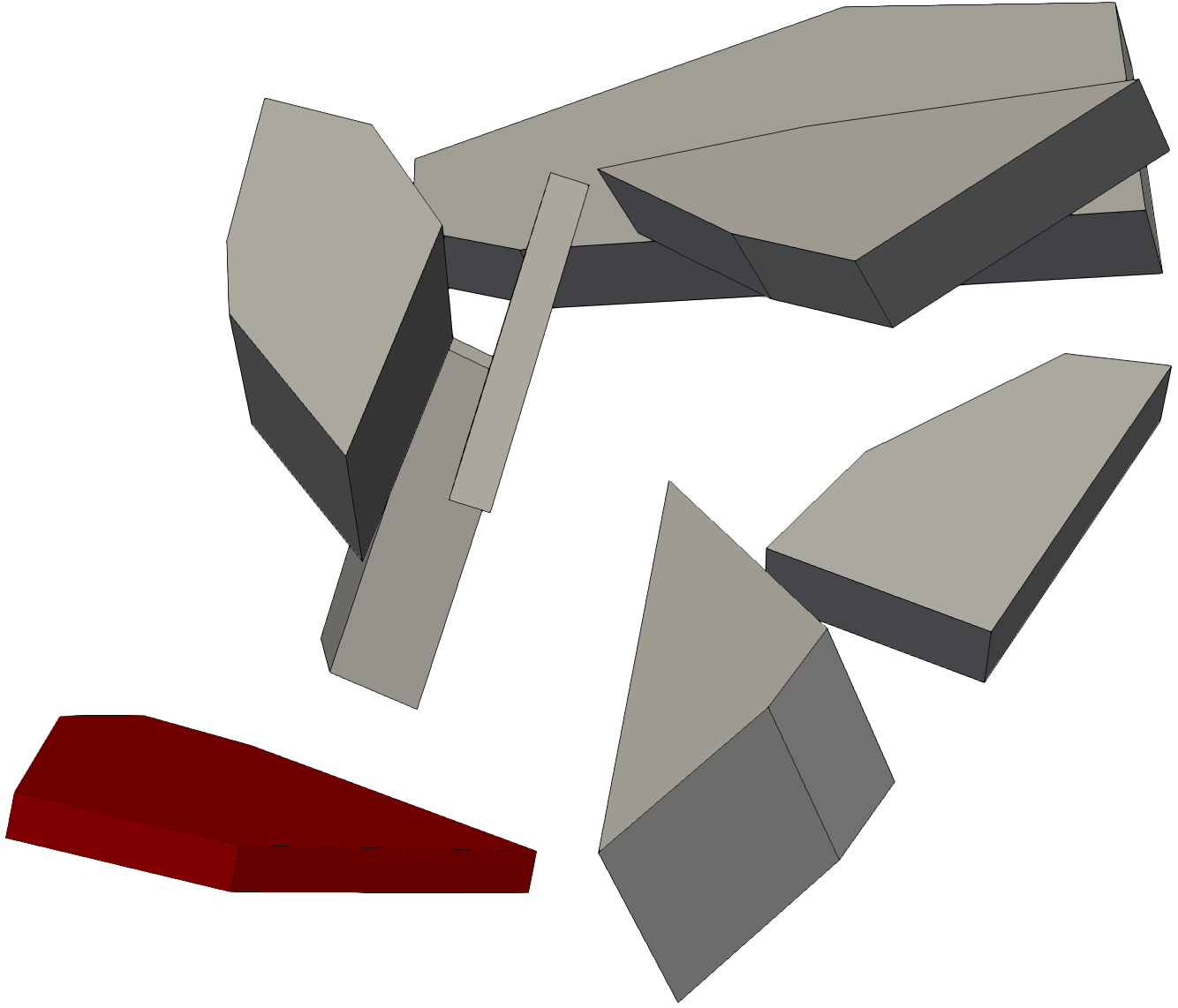}
 \caption{$t=1$}
\end{subfigure}
\end{center}
\caption{We define polyhedra geometries by extruding the polygons in Figure \ref{fig:polygon-contact} in the $z$ direction---the height of each polygon is randomly choosen from $\{0.1, 0.25, 0.5\}$. We use the same viscoelastic contact law and prescribe an initial velocity of $v_0 = (0.25, 0.25, 0)$ $m/s$ to the red particle.}
\label{fig:polyhedra-contact}
\end{figure}

ParticLS can also simulate more complicated geometries. For example, convex polyhedra, which generalizes the box geometry (see Figure \ref{fig:particle-geometry}), are defined by a set of faces outlining the shape's perimeter---lines in two dimensions and polygons in there dimensions. The signed distance for a polyhedra with $m$ faces is determined by calculating the minimum distance $S_i$ between the point $X$ and the $i^{\mbox{th}}$ face. The SDF for a polygon is therefore
\begin{equation}
    \varphi(X) = \min_{i \in \{1,\hdots,m\}}{(S_i)}.
    \label{eq:SDF-poly}
\end{equation}
Figures \ref{fig:polygon-contact} and \ref{fig:polyhedra-contact} show irregularly shaped polygons in two dimensions, and polyhedra in three dimensions, respectively. 

\subsection{Time-dependent particle geometries}

ParticLS allows users to create objects with time-dependent properties by implementing children of the \texttt{StateObject} class. In this example, we simulate a dilating sphere. We implement a class \texttt{SphereRadius}, which is a one dimensional child of \texttt{StateObject} whose right hand side defines how quickly the sphere's radius changes as a function of time. In this example, the radius simply increases linearly with respect to time. However, users can implement more complex models that change the particle's radius according to a function $dR(t)/dt = g(R(t), t; \theta)$, where $g$ is a generic function, and $\theta$ parameterizes a model that could simulate thermal expansion or a change in particle shape due to other material properties. 

\begin{figure}[ht!]
\captionsetup[subfigure]{labelformat=empty}
\begin{center}
\begin{subfigure}[b]{0.45\linewidth}
    \includegraphics[width=\linewidth]{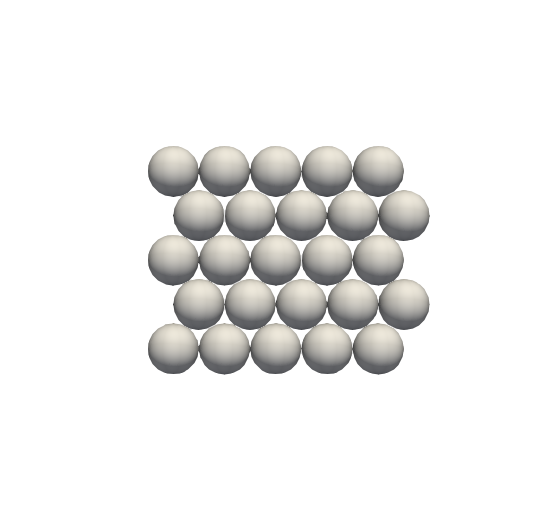}
    \caption{t=1}
\end{subfigure}
\begin{subfigure}[b]{0.45\linewidth}
    \includegraphics[width=\linewidth]{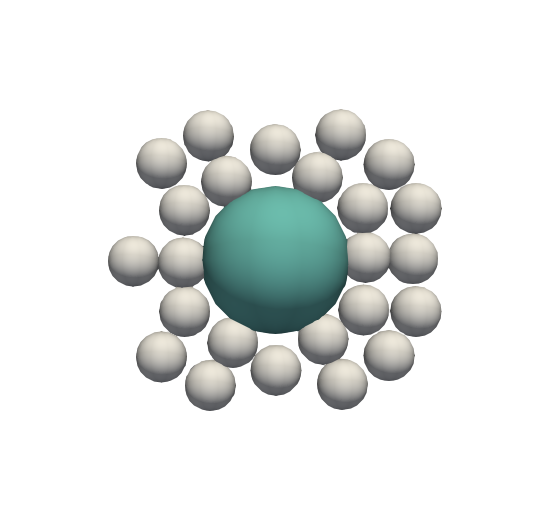}
    \caption{t=2}
\end{subfigure}
\begin{subfigure}[b]{0.45\linewidth}
    \includegraphics[width=\linewidth]{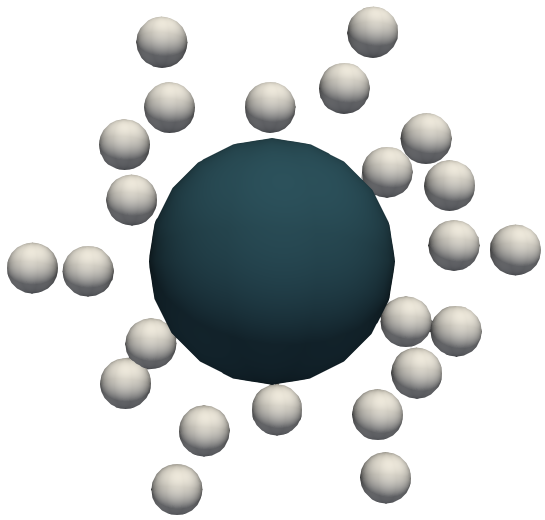}
    \caption{t=3}
\end{subfigure}
\begin{subfigure}[b]{0.45\linewidth}
    \includegraphics[width=\linewidth]{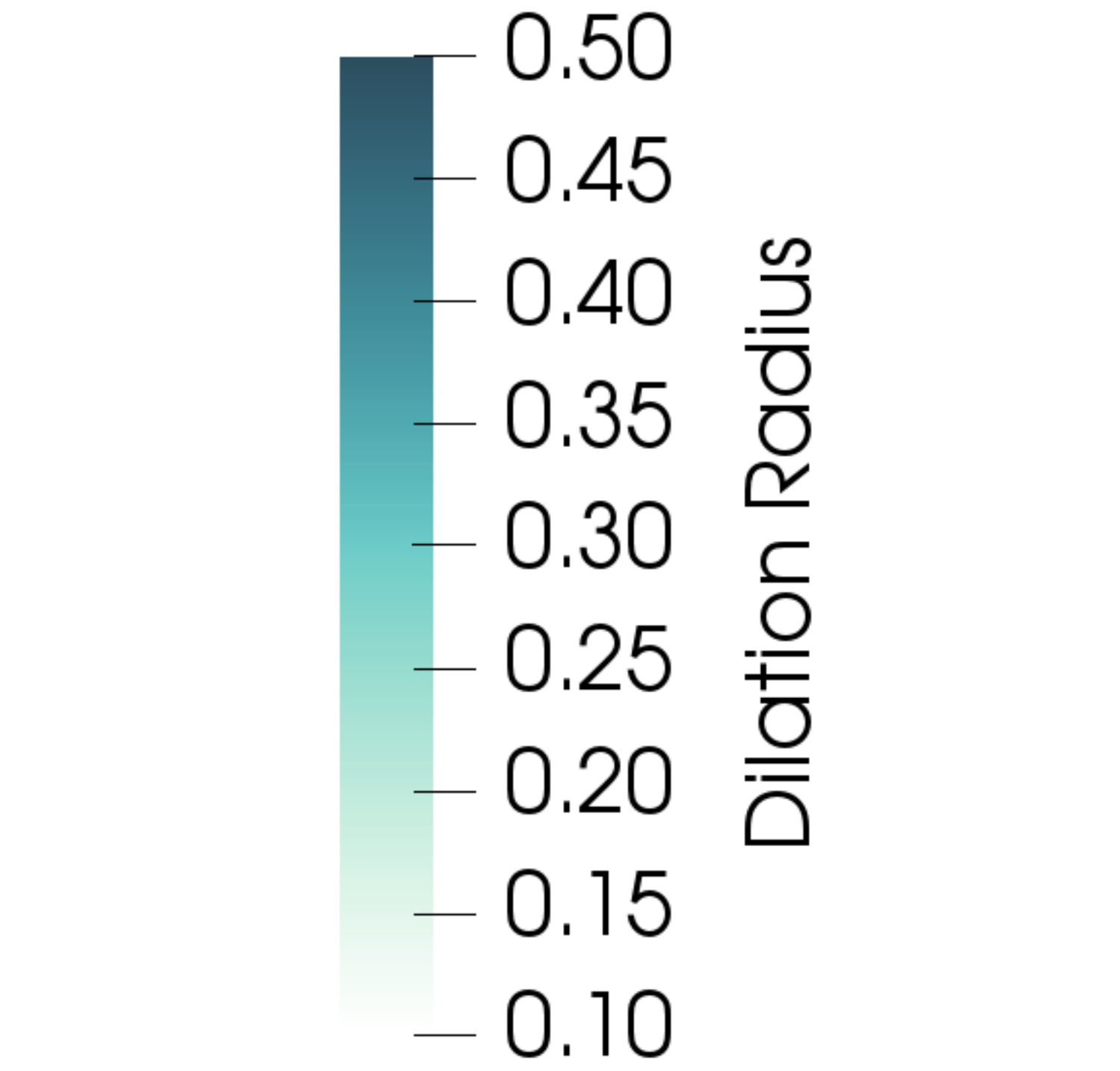}
\end{subfigure}
\end{center}
\caption{Example of a particle with a time-dependent geometry. Initially each of the hex packed particles are the same size but the radius of the middle particle grows according to $dR(t)/dt=1$ $m/s$.}
\label{fig:time-dependent}
\end{figure}

We demonstrate this time-dependent functionality with a simple example in which normal sphere particles are displaced by one that expands. We begin with $25$ equally-sized spheres that are configured in a hexagonal packing configuration with $5$ rows of $5$ spheres (Figure \ref{fig:time-dependent}). The radius of each sphere is defined by a \texttt{SphereRadius} object where $dR(t)/dt = 0$ $m/s$, except the middle sphere where $dR(t)/dt = 1$ $m/s$. In addition to adding each \texttt{RigidBody} to the \texttt{Collection} object, we also add the \texttt{SphereRadius} with non-zero right hand side. The \texttt{VelocityVerlet} time-integrator evolves both the particles and the radius in time. The contact law between objects is the elastic contact described in section \ref{sec:elastic-contact} with with parameters $\kappa_n = 20$ $N/m$, $\kappa_s = \kappa_n/2$, and $\mu = 0.3$. The results in Figure \ref{fig:time-dependent} show how this time-dependent capability gives users one approach to infuse additional physics into their ParticLS simulation beyond the rigid body mechanics provided with DEM.

\subsection{Multiscale granular media mechanics}

ParticLS tracks micro-scale particles (e.g. soil grains), whose aggregate properties determine the macro-scale response of the granular material (e.g. soil strength). In general, the DEM is a useful tool for modeling the mechanical response of granular media and studies have been able to match modeled results with laboratory experiments \cite{Kawamotoetal2016,Mcdowell2006,Shmulevich2007,Asaf2007}. However, large domains with millions or billions of particles can be prohibitively expensive. One approach for modeling these large systems is with multiscale methods, of which there are a few different options. One method, called the hierarchical multiscale method (HMM), leverages the finite element method (FEM) over the full domain and uses small-scale DEM simulations as local representative volume elements (RVEs) to capture micro-scale dynamics where necessary \cite{GuoZhao2014}. In this example, we present the general principles of the HMM approach and show how the results of a ParticLS DEM simulation can fit into the framework.

A classic problem for illustrating the benefit of the HMM is simulating the response of granular media under different loading conditions. In this case we are concerned with relating the applied loads to the material's resultant deformation. In a normal FEM approach to this problem, the governing equation can be described as:
\begin{equation}
    \sigma = D \varepsilon
    \label{eq:multiscale-gov-eqn}
\end{equation}
where $\sigma$ is the stress, $D$ is the tangent operator, and $\varepsilon$ is the strain \cite{GuoZhao2015}. Unfortunately, this requires a constitutive law to define $D$ beforehand, which introduces assumptions and biases regarding the material's response. However, the HMM approach avoids this issue by explicitly solving for the stress in the granular material by simulating a DEM particle collection with the given strain boundary conditions. In the HMM, information is transferred between the two models in an iterative manner such that the FEM solution provides strain boundary conditions ($\varepsilon$) to each DEM simulation, and the resolved DEM solution returns stress ($\sigma$) and tangent operator ($D$) to the FEM \cite{GuoZhao2015}. Therefore, we can solve the equation \ref{eq:multiscale-gov-eqn} without making any assumptions about the material's constitutive response. 

The stress in the DEM particle collection is calculated with the average Cauchy stress tensor \eqref{eqn:cauchy-stress}, and the tangent operator is calculated with 
\begin{equation}
    D = \frac{1}{V}\sum_{N_{c}}(k_n \boldsymbol{n^{c}} \otimes \boldsymbol{d^{c}} \otimes \boldsymbol{n^{c}} \otimes \boldsymbol{d^{c}} + k_t \boldsymbol{t^{c}} \otimes \boldsymbol{d^{c}} \otimes \boldsymbol{t^{c}} \otimes \boldsymbol{d^{c}})
    \label{eq:multiscale-tan-op}
\end{equation}
where $k_n$ and $k_t$ are the normal and tangential stiffness, $\boldsymbol{n^{c}}$ and $\boldsymbol{t^{c}}$ are the normal and tangential directions of each particle contact, and $\boldsymbol{d^{c}}$ is the vector between each pair of particles in contact. This feedback loop between DEM and FEM solvers is repeated every time the FEM model requires fine scale information.

\begin{figure}
\centering
\begin{minipage}{.5\linewidth}
  \centering
  \includegraphics[width=\linewidth]{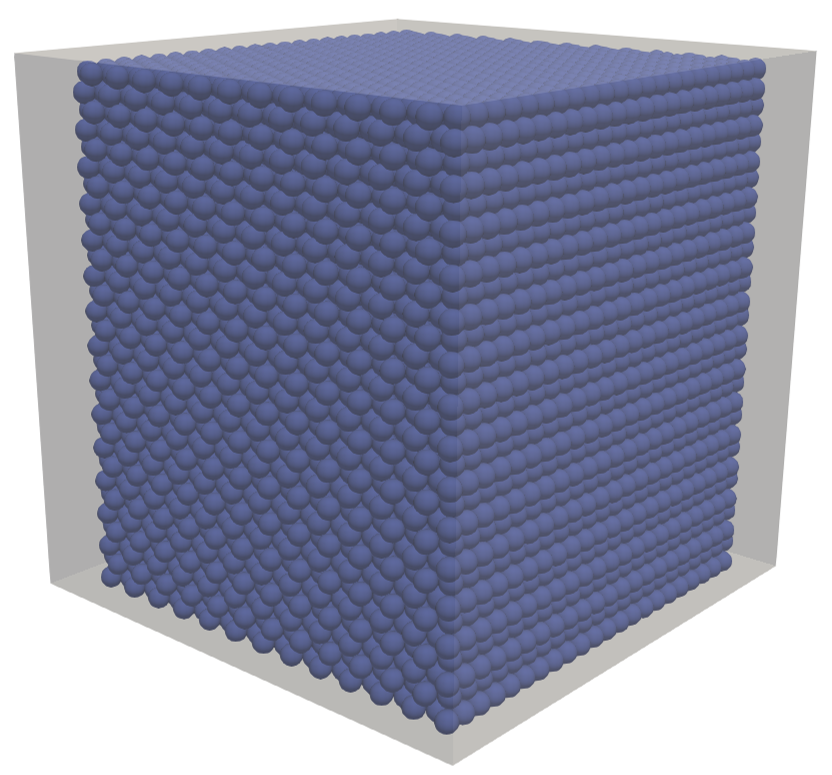}
\end{minipage}%
\begin{minipage}{.5\linewidth}
  \centering
  \includegraphics[width=\linewidth]{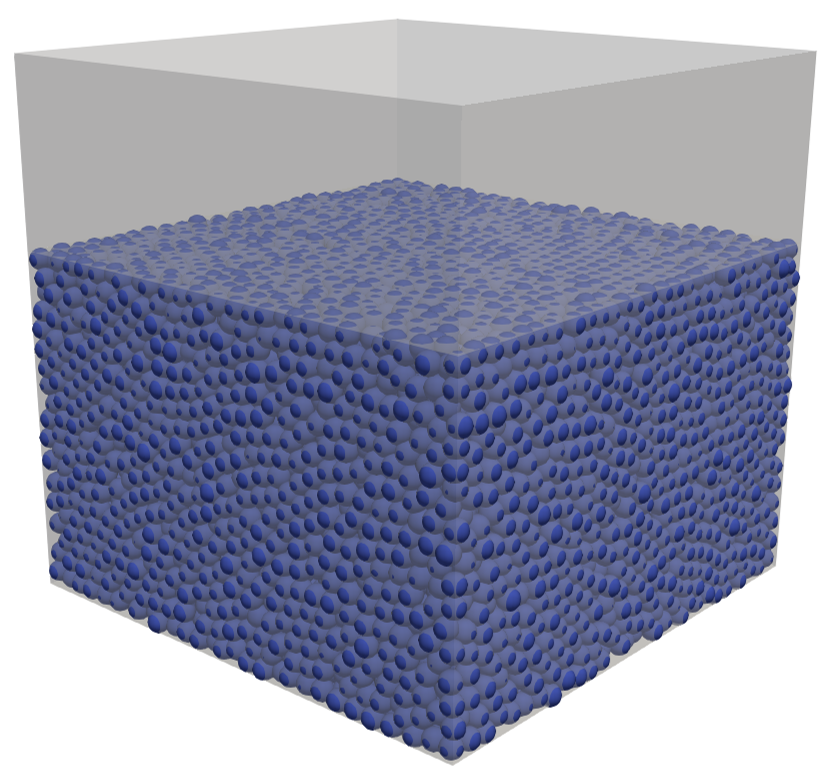}
\end{minipage}
\caption{Example of uniaxial compression of a sphere particle assembly with 16500 particles and initial dimensions of $0.85\times0.85\times1.00$.}
\label{fig:granular-media}
\end{figure}

In this example, we show how ParticLS' DEM capabilities can be used to simulate an RVE in uniaxial compression---in HMM this strain would come from the FEM solution. We track the average Cauchy stress tensor of a confined particle assembly as it is compressed in the $z$ direction. The RVE consists of $16500$ hex-packed spheres that start in a $0.85\times0.85\times1.00$ box. The sides and bottom of the box are planar particles whose location and orientation are fixed in time. The top of the box is another planar particle whose velocity is fixed at $v = (0, 0, -0.35)$ $m/s$. 
We run the simulation for $t=1$ units of time with a viscoelastic contact model with parameters $\kappa_n = 20$ $N/m$, $\kappa_s = \kappa_n/2$, $\gamma_n = 0.08$ $Ns/m$, $\gamma_s = \gamma_n/2$, and $\mu = 0.3$; Figure \ref{fig:granular-media} shows the initial and final configurations. A \texttt{CollectionTracker} was used to track the average Cauchy stress tensor as a function of time.

\begin{figure}[ht!]
\begin{center}
\begin{tikzpicture}
    \begin{axis}[width=0.9\linewidth, height=0.75*\axisdefaultheight,
                xlabel=Strain $\varepsilon$, ylabel=Stress $\sigma_{zz}$, 
                ymin=-45, ymax=5, xmin=0, xmax=0.35,
                xtick pos=left, ytick pos=left, axis lines=left,
                y tick label style={/pgf/number format/.cd, fixed, fixed zerofill, precision=1, /tikz/.cd},
                x tick label style={/pgf/number format/.cd, fixed, fixed zerofill, precision=2, /tikz/.cd}]
    \draw [gray,dashed,thick] (0,450) -- (350,450);
    \addplot[color=darkgray, very thick] plot coordinates {
            (0.0, -4.55856e-13)
            (0.0035, -0.0259875)
            (0.007, -0.04095)
            (0.0105, -0.0448875)
            (0.014, -1.31069)
            (0.0175, -2.35273)
            (0.021, -2.8521)
            (0.0245, -2.94102)
            (0.028, -2.65506)
            (0.0315, -2.0431)
            (0.035, -1.25137)
            (0.0385, -0.407328)
            (0.042, 0.213662)
            (0.0455, 0.282529)
            (0.049, -0.113715)
            (0.0525, -0.315906)
            (0.056, -1.26785)
            (0.0595, -1.4038)
            (0.063, -1.46901)
            (0.0665, -1.51873)
            (0.07, -1.58042)
            (0.0735, -1.62)
            (0.077, -1.63081)
            (0.0805, -1.67247)
            (0.084, -1.7726)
            (0.0875, -1.84772)
            (0.091, -1.93402)
            (0.0945, -1.98151)
            (0.098, -3.10094)
            (0.1015, -3.26657)
            (0.105, -3.51107)
            (0.1085, -3.75224)
            (0.112, -4.03391)
            (0.1155, -4.18242)
            (0.119, -4.35675)
            (0.1225, -4.63976)
            (0.126, -4.91748)
            (0.1295, -5.18597)
            (0.133, -5.3553)
            (0.1365, -6.52005)
            (0.14, -6.9082)
            (0.1435, -7.2247)
            (0.147, -7.58372)
            (0.1505, -7.9569)
            (0.154, -8.35046)
            (0.1575, -8.67976)
            (0.161, -9.09738)
            (0.1645, -9.41808)
            (0.168, -9.74906)
            (0.1715, -10.0706)
            (0.175, -11.2798)
            (0.1785, -11.7259)
            (0.182, -12.07)
            (0.1855, -12.4244)
            (0.189, -12.8145)
            (0.1925, -13.2788)
            (0.196, -13.6731)
            (0.1995, -14.0735)
            (0.203, -14.4573)
            (0.2065, -14.8237)
            (0.21, -15.2284)
            (0.2135, -15.6039)
            (0.217, -17.0331)
            (0.2205, -17.5539)
            (0.224, -17.9357)
            (0.2275, -18.3345)
            (0.231, -18.7539)
            (0.2345, -19.2597)
            (0.238, -19.5711)
            (0.2415, -20.0575)
            (0.245, -20.6712)
            (0.2485, -21.2087)
            (0.252, -21.6007)
            (0.2555, -22.8277)
            (0.259, -23.467)
            (0.2625, -24.0914)
            (0.266, -24.5928)
            (0.2695, -25.2386)
            (0.273, -25.7727)
            (0.2765, -26.2211)
            (0.28, -26.7151)
            (0.2835, -27.2901)
            (0.287, -27.7508)
            (0.2905, -28.2099)
            (0.294, -28.67)
            (0.2975, -29.6822)
            (0.301, -30.6351)
            (0.3045, -31.3426)
            (0.308, -31.7395)
            (0.3115, -32.3673)
            (0.315, -33.061)
            (0.3185, -33.7244)
            (0.322, -34.1398)
            (0.3255, -34.6198)
            (0.329, -35.2979)
            (0.3325, -35.8807)
            (0.336, -37.2268)
            (0.3395, -38.0451)
            (0.343, -38.2278)
(0.3465, -39.4941)};
    \end{axis}
\end{tikzpicture}
\end{center}
\caption{Stress-strain plot for uniaxial compression experiment in Figure \ref{fig:granular-media}, where the plotted stress value is the $\sigma_{zz}$ component of \eqref{eqn:cauchy-stress} and the strain is estimated as the change in position of the top plane.}
\label{fig:stress-strain}
\end{figure}
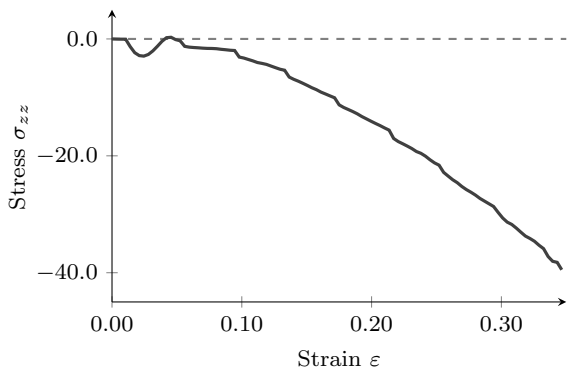

This example illustrates ParticLS' ability to track macro-scale properties that are important when modeling granular media---specifically, how ParticLS fits into a multiscale framework. The \texttt{CollectionTracker} measures the collection's average Cauchy stress and in this case a \texttt{ParticleTracker} estimates the strain as the change in top plane's position. 
The collection's stress-strain response throughout the experiment is illustrated in Figure \ref{fig:stress-strain}, which plots the $\sigma_{zz}$ component of the stress against the strain. The stress-strain response oscillates at the beginning as the top plate meets the top row of particles and they spread out to fill the box domain. After initial contact, the stress increases as the collection is compressed. The example RVE simulation (Figures \ref{fig:granular-media} and \ref{fig:stress-strain}) shows how ParticLS could be used to model the RVE's used in an HMM multi-scale framework for different geomechanic systems such as soil, snow, or sea ice \cite{GuoZhao2014,GuoZhao2016,Bobillier12018,Gaume2014,Hagenmuller2015}. 

\subsection{Peridynamic model of ordinary material} \label{sec:peridynmic-plate}

\sloppy We demonstrate ParticLS' ability to model peridynamic solids by stretching a plate that is initially defined by the material domain $\Omega = 1 \times 1 \times 0.01$ plate. We denote the components of $x \in \Omega$ by $x = (x_0, x_1, x_2)$. The horizon is determined by a ball with radius $\delta$, $\bar{\mathcal{B}}_{\delta}(x) = \mathcal{B}_{\delta}(x) \cap \Omega$---each material point $x \in \Omega$ interacts with the points $x^{\prime}$ such that $\|x-x^{\prime}\|<\delta$ and $x^{\prime} \in \Omega$. We consider a linear peridynamic solid \cite{Sillingetal2007} with a hole cut through it, that fractures given a driving force on its right boundary.

We implement a linear peridynamic solid---refer to sections 13-15 of \cite{Sillingetal2007} for additional details. Given $x \in \Omega$ and $x^{\prime} \in \bar{\mathcal{B}}_{\delta}(x)$, define the extension
\begin{equation}
    e(x, x^{\prime}, t) = \|(x^{\prime}+u(x^{\prime}, t)) - (x+u(x, t))\|-\|x^{\prime} - x\|,
\end{equation}
which is a proxy for the strain in the bond between $x$ and $x^{\prime}$. Additionally, define the weighted volume 
\begin{equation}
    m(x) = \int_{\bar{\mathcal{B}}_{\delta}(x)} \omega(\|x^{\prime}-x\|) \|x^{\prime}-x\|^2 \, dx^{\prime}
\end{equation}
and the dilation parameter 
\begin{equation}
    \theta(x, t) = \int_{\bar{\mathcal{B}}_{\delta}(x)} \omega(\|x^{\prime}-x\|) \|x^{\prime}-x\| e(x, x^{\prime}, t) \, dx^{\prime}, 
\end{equation}
where $\omega: \mathbb{R}^{+} \mapsto \mathbb{R}^{+}$ is the influence function. We choose 
\begin{equation}
    \omega(\|\xi\|) = \exp{(-\|\xi\|^2/\delta^2)}.
\end{equation}
The isotropic and deviatoric parts of $e(x, x^{\prime}, t)$ are 
\begin{subequations}
\begin{eqnarray}
 e^{i}(x, x^{\prime}, t) &=& \theta(x, t) \|x^{\prime}-x\|/3 \\
 e^{d}(x, x^{\prime}, t) &=& e(x, x^{\prime}, t) - e^{i}(x, x^{\prime}, t),
\end{eqnarray}
\end{subequations}
respectively. A linear peridynamic solid is defined by the force state \eqref{eq:ordinary-force-state}
\begin{equation}
    \begin{split}
        f(x, x^{\prime}, t) = & \omega(\|x^{\prime}-x\|) (3 k \theta(x, t) \|x^{\prime}-x\| \\ & - 15 \mu e^{d}(x, x^{\prime}, t)) / m(x),
    \end{split}
    \label{eq:linear-peridynamic-solid}
\end{equation}
where $k>0$ and $\mu \geq 0$ are the bulk and shear moduli. The velocities at $x=0 $ and $x=1$ are zero in the $y$ and $z$ directions, but we apply a velocity of $0.1\text{/sec}$ in the $x$ direction.

\begin{figure}
    \centering
    \begin{subfigure}[b]{\linewidth}
        \centering
        \includegraphics[width=0.4\linewidth]{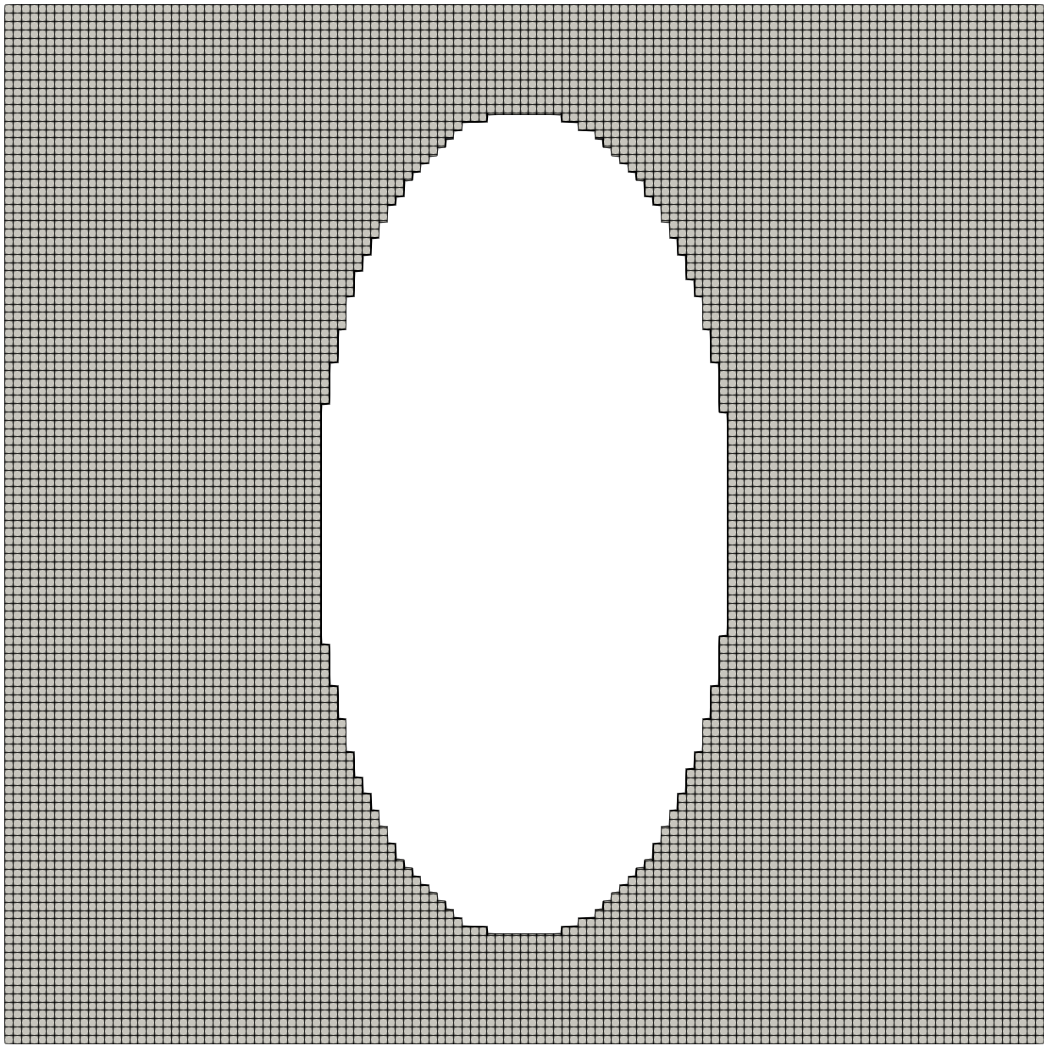}%
        \caption{$t=0s$}
    \end{subfigure}
    \vskip\baselineskip
    \begin{subfigure}[b]{\linewidth}
        \centering
        \includegraphics[width=0.47\linewidth]{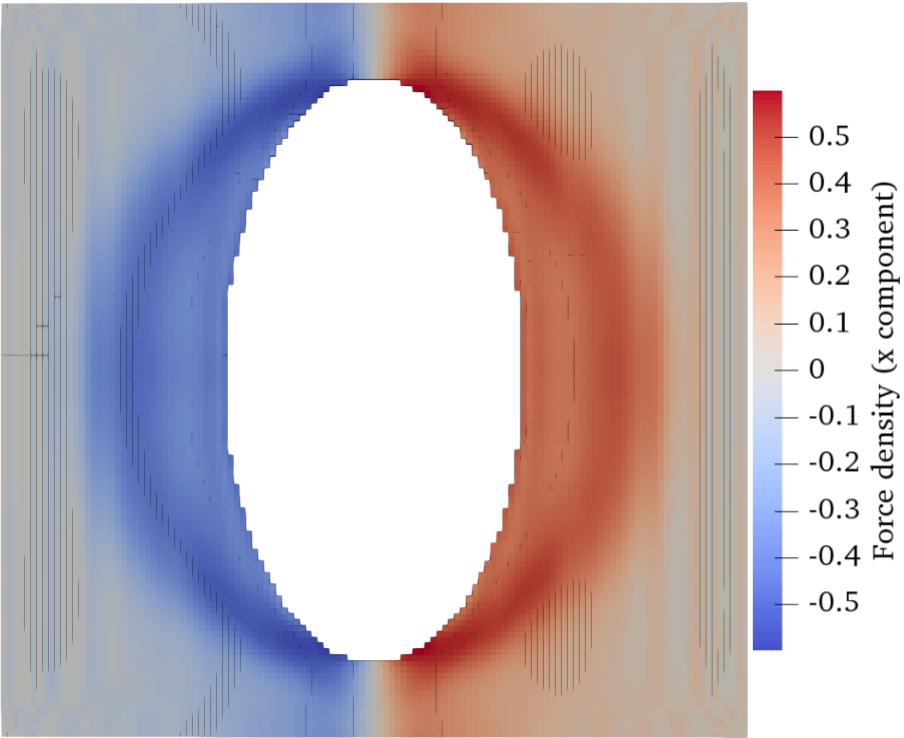}%
        \hspace{4mm}
        \includegraphics[width=0.47\linewidth]{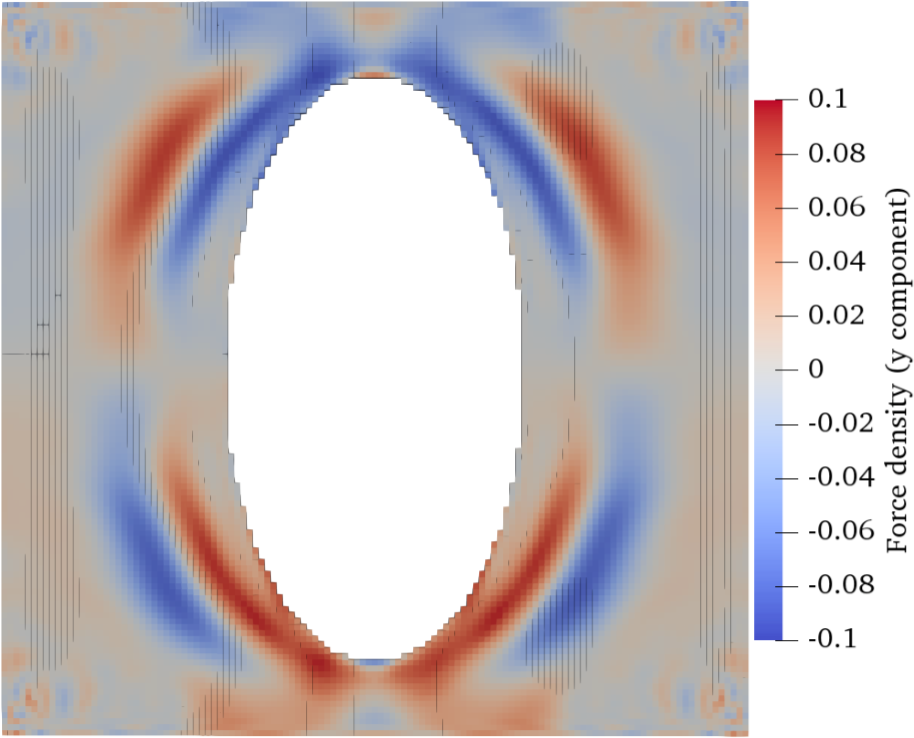}
        \caption{$t=0.25s$}
    \end{subfigure}
    \vskip\baselineskip
    \begin{subfigure}[b]{\linewidth}
        \centering
        \includegraphics[width=0.47\linewidth]{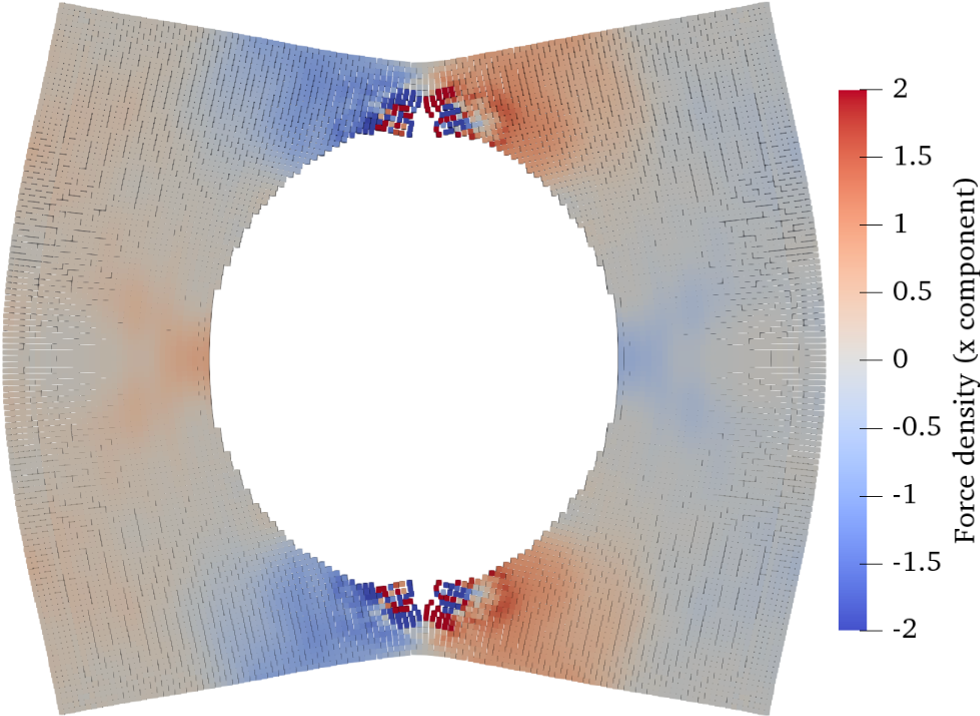}%
        \hspace{4mm}
        \includegraphics[width=0.47\linewidth]{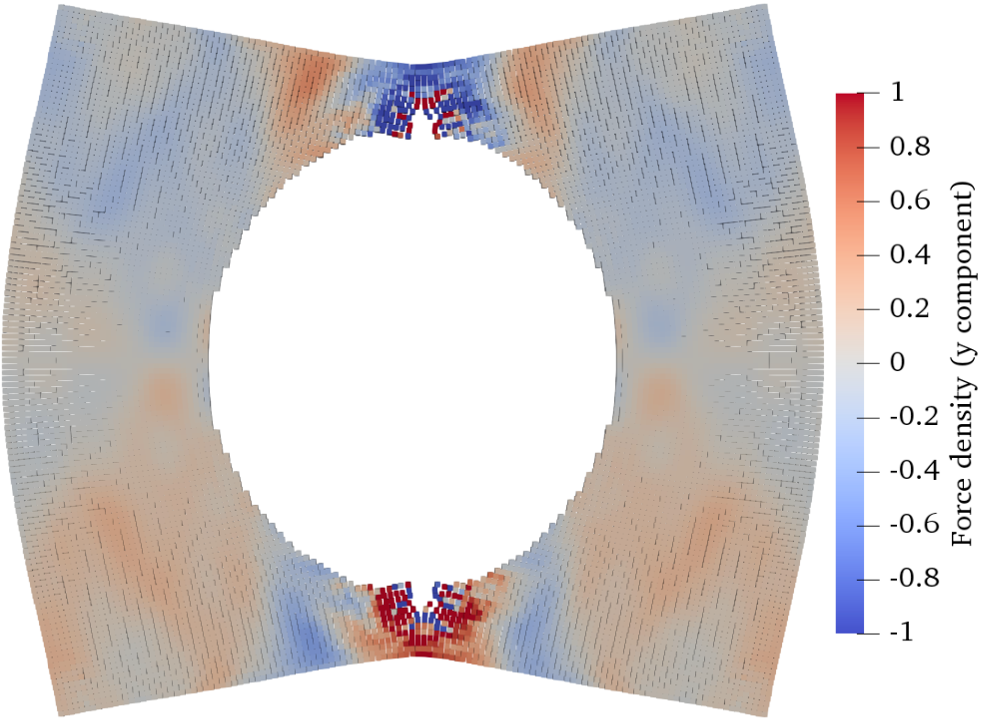}
        \caption{$t=1.0s$}
    \end{subfigure}
    \caption{Results of a linear peridynamic plate with a hole that is stretched to the right under a driving force. The left and right columns of (b) and (c) are the $x$ and $y$ components of the force density, respectively.}
    \label{fig:plate-with-hole}
\end{figure}

We discretize the domain $\Omega$ using $1/N_x \times 1/N_y \times 1/N_z$ cubes whose SDF is defined by \eqref{eq:SDF-box}. ParticLS integrates the discrete peridynamic equation \eqref{eq:discrete-peridynamic-equation} with respect to time. We cut a hole through the plate by removing all particles whose initial center of mass $\bar{x}$ satisfies 
\begin{equation}
    (\bar{x}_1-0.5)^2/r_1^2 + (\bar{x}_2-0.5)^2/r_2^2 < 1,
\end{equation}
where $\bar{x}_1$ and $\bar{x}_2$ are the first and second components of $\bar{x}$ and $r_1$ and $r_2$ are radii of an ellipse in the $x$ and $y$ directions, respectively. Figure \ref{fig:plate-with-hole} shows the initial location of the particles with $N_x = N_y = 125$ and $N_z = 1$. We model internal forces using the linear peridynamic solid force state defined by \eqref{eq:linear-peridynamic-solid}. However, we drive the simulation by prescribing a body force 
\begin{equation}
    \beta(x, t) = \begin{cases}
        3(1-\exp{(-10 t)}) & \mbox{if } x>0.95 \\
        -3(1-\exp{(-10 t)}) & \mbox{if } x<0.05 \\
        0 & \mbox{else}
    \end{cases}
\end{equation}
on the portion of the plate such that $\bar{x}_1>0.05$. In Figure \ref{fig:plate-with-hole}, we show results for our plate-with-hole experiment. The plate deforms most at the top and bottom boundaries of the hole, as we would expect, and fractures nucleate in these regions due to the stress concentrations. Traditional continuum mechanic models---which define a differential constitutive law---generate a singularity at crack tip, which makes modeling with these methods changeling, if not impossible. The peridynamic framework in ParticLS allows us to model deforming and fracturing bodies. After fracture, the resulting bodies interact using DEM contact laws, which allows us to model materials that break into several discrete pieces within one simulation. 
{\color{EDIT_COLOR}{In Figure \ref{fig:plate-with-hole}, we visualize the material points in the plate using cubes. In the peridynamic setting, this geometric choice is irrelevant: the peridynamic law only requires the distance between the center-of-masses. However, after fracture occurs, we replace the contact law with elastic contact allowing the resulting halves of the plate to interact as discrete, deforming bodies. Therefore, the chosen particle geometry only matters \emph{after} fracture.}}

\subsection{Peridynamic model of a cantilevered beam}

To illustrate ParticLS' ability to match analytical solutions for continuum mechanic problems, we simulate the deflection of a cantilevered beam in two and three dimensions. We discretize the beam using material points---similar to Section \ref{sec:peridynmic-plate}---and connect nearby points (within radius $r$) with a beam that models internal forces. As the particles move, the beam deforms (compresses, stretches, and bends), exerting a material force on the connected particles following Euler-Bernoulli beam theor---\cite{Andreetal2012CohesiveBeam} derive the corresponding expression for the force and torque that this exerts on the material points. 

We propose that this beam model is akin to a peridynamic model. As the number of particles $n$ discretizing the beam increases, each particle responds to deformation to the material within a radius $r$. We let $r \rightarrow 0$ as $n \rightarrow \infty$---\cite{SillingLehoucq2008} to show that in this limit the peridynamic model converges to classical elastic theory. We, therefore, compare the deflection of the discretized beam with bonded adjacent particles to the classical analytic solution for a cantilevered beam bending due to an external acceleration. There are two key differences between this simulation and how we typically define a peridynamic model. First, we have not explicitly defined the force state $F[x, x^{\prime}, t]$ that is used in Equation \eqref{eq:peridynamic-force-density}. Second, the beams impose a torque on the material point, which requires us to evolve the material point rotation and angular velocity in order to conserve angular momentum. Traditional peridynamics---even state-based peridynamics---assumes angular momentum is conserved by imposing requirements on the force state. We propose this as a peridynamic-like model but exploring how to formally include angular momentum in the peridynamic framework is beyond the scope of the current paper. For our purposes, we numerically validate the model and our code by matching an analytic solution and showing expected dynamic behavior.

\subsubsection{Two dimensional case}

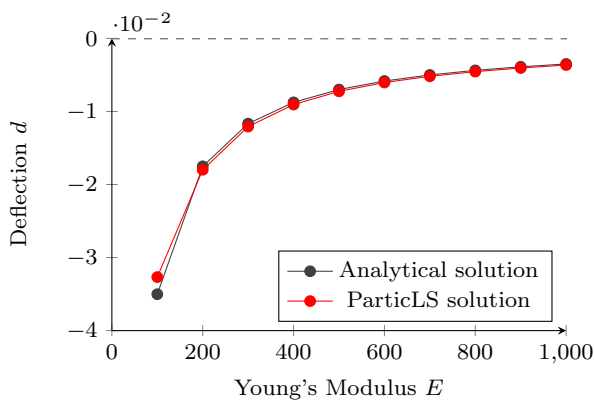
\begin{figure}[ht!]
\begin{center}
\begin{tikzpicture}
    \begin{axis}[width=0.9\linewidth, height=0.75*\axisdefaultheight,
                xlabel=Young's Modulus $E$, ylabel=Deflection $d$, 
                ymin=-0.04, ymax=0.0, xmin=0.0, xmax=1000.0,
                xtick pos=left, ytick pos=left, axis lines=left,
                y tick label style={/pgf/number format/.cd, fixed, fixed zerofill, precision=0, /tikz/.cd},
                x tick label style={/pgf/number format/.cd, fixed, fixed zerofill, precision=0, /tikz/.cd}, legend pos=south east]
    \draw [gray,dashed,thick] (0,400) -- (1000,400);
    \addplot[color=darkgray, mark=*] plot coordinates {
            (100.0, -0.03501409)
            (200.0, -0.01750705)
            (300.0, -0.01167136)
            (400.0, -0.00875352)
            (500.0, -0.00700282)
            (600.0, -0.00583568)
            (700.0, -0.00500201)
            (800.0, -0.00437676)
            (900.0, -0.00389045)
            (1000.0, -0.00350141)};
    \addlegendentry{Analytical solution}
    \addplot[color=red, mark=*] plot coordinates {
            (100.0, -0.0326745)
            (200.0, -0.017988)
            (300.0, -0.0120749)
            (400.0, -0.00906304)
            (500.0, -0.00725115)
            (600.0, -0.00604474)
            (700.0, -0.00519167)
            (800.0, -0.00455208)
            (900.0, -0.00405583)
            (1000.0, -0.00364539)};
    \addlegendentry{ParticLS solution}
    \end{axis}
\end{tikzpicture}
\end{center}
\caption{Comparison of numerical beam deflection to analytical solution for a constant external acceleration of $a=-2E-6$ $m/s^{2}$. We set the physical parameters $M=1$ $kg$, $L=1$ $m$, and $I=7.85E-9$ $m^{4}$.}
\label{fig:analytical-beam}
\end{figure}

The analytical solution for the maximum deflection, $d$, of a cantilevered beam acting under an external acceleration, $a$, is as follows:
\begin{equation}
    d = \frac{M a L^{4}}{8 E I}.
\end{equation}
where $M$ is the beam mass, $L$ is the beam length, and $I$ is the beam moment of inertia. Figure \ref{fig:analytical-beam} shows the comparison between $n=11$ material points along the beam and the expected analytical result. It is obvious from this result that ParticLS is able to accurately approximate the deflection of a continuous beam.

\subsubsection{Three dimensional case}

\begin{figure}
    \centering
    \begin{subfigure}[b]{\linewidth}
        \centering
        \includegraphics[width=0.4\linewidth]{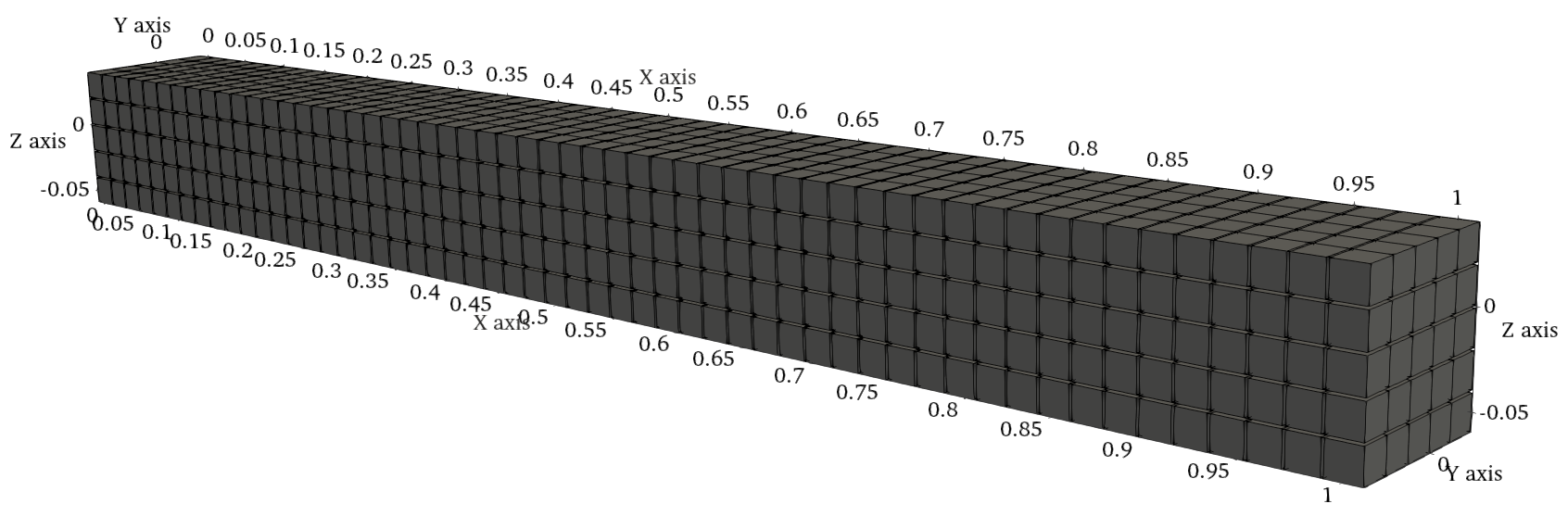}%
        \caption{$t=0$}
    \end{subfigure}
    \vskip\baselineskip
    \begin{subfigure}[b]{\linewidth}
        \centering
        \includegraphics[width=0.47\linewidth]{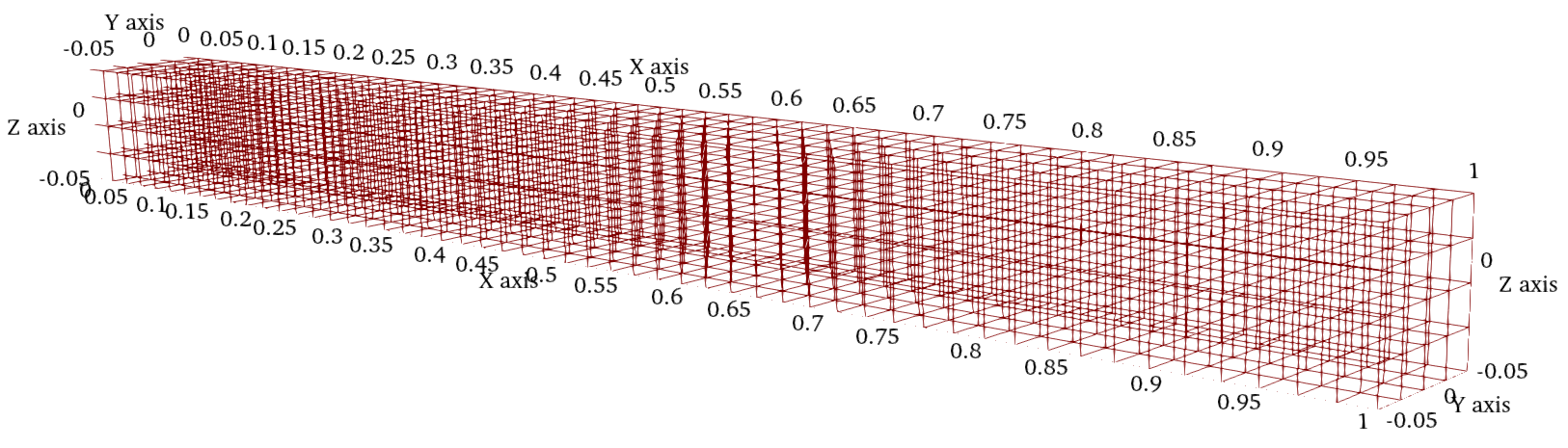}%
        \hspace{4mm}
        \includegraphics[width=0.47\linewidth]{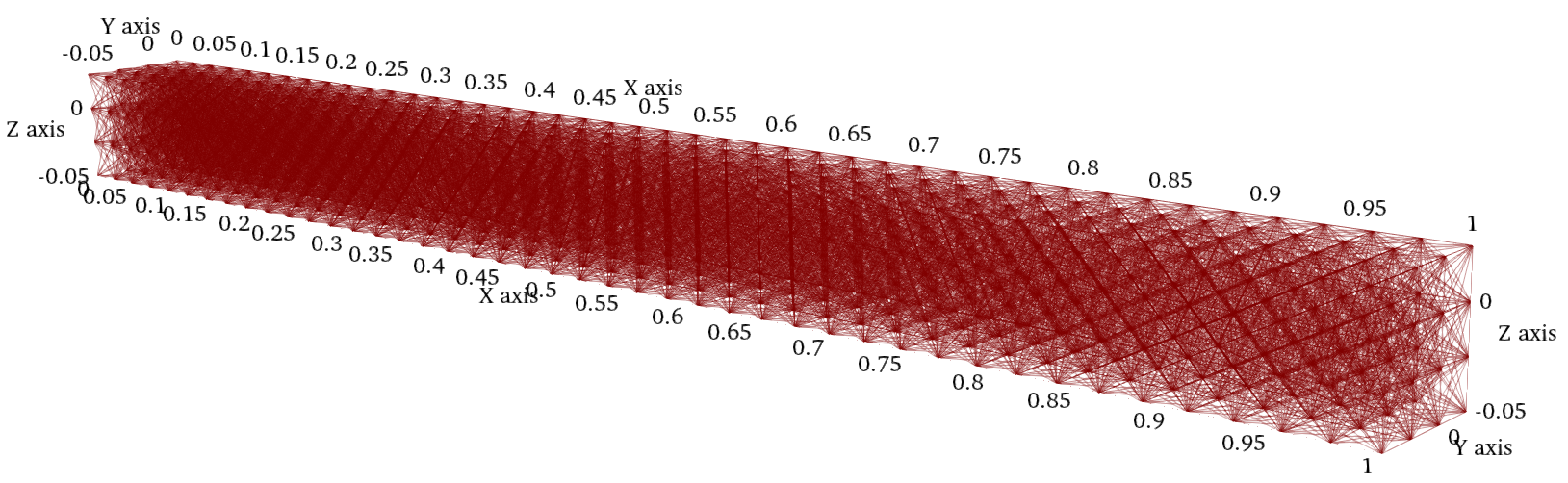}
        \caption{Material point bonds}
    \end{subfigure}
    \vskip\baselineskip
    \begin{subfigure}[b]{\linewidth}
        \centering
        \includegraphics[width=0.47\linewidth]{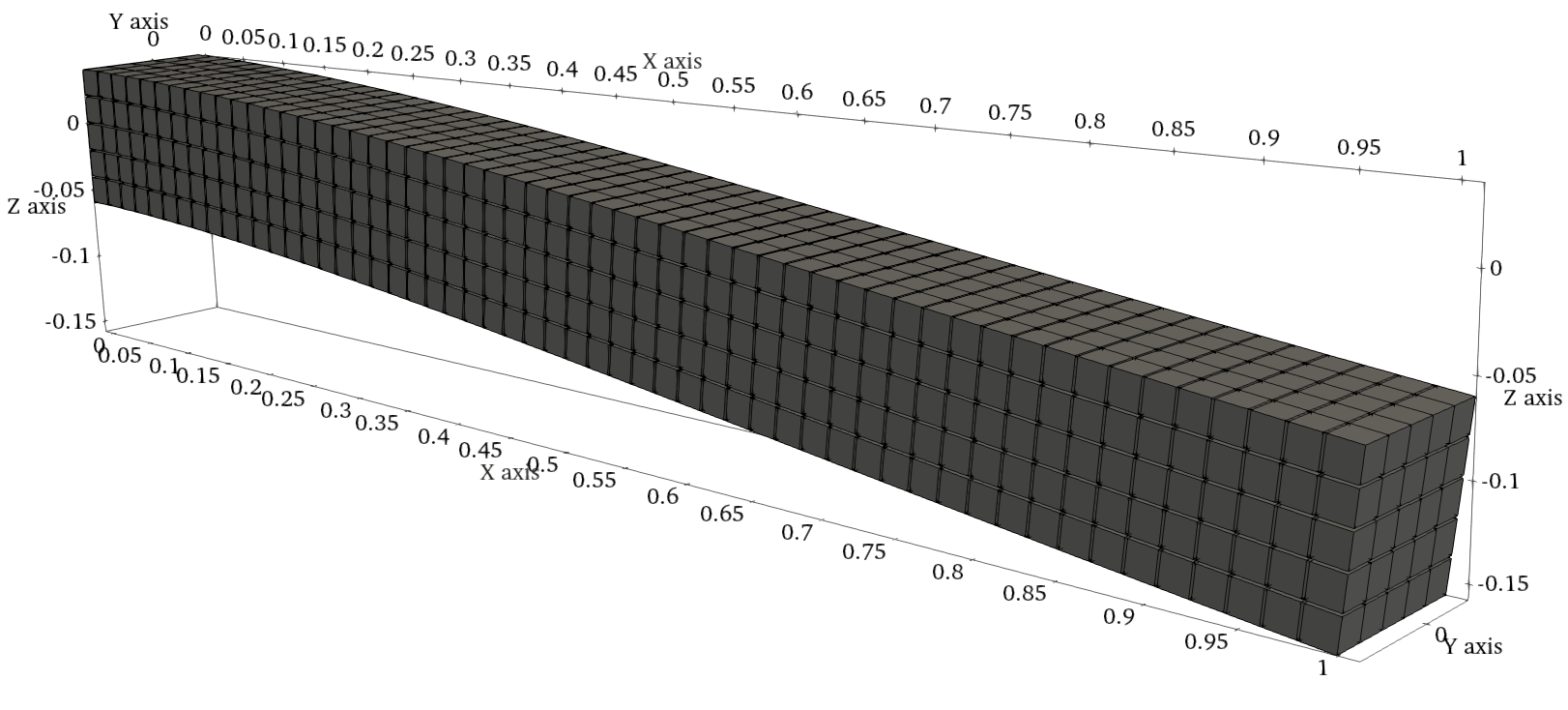}%
        \hspace{4mm}
        \includegraphics[width=0.47\linewidth]{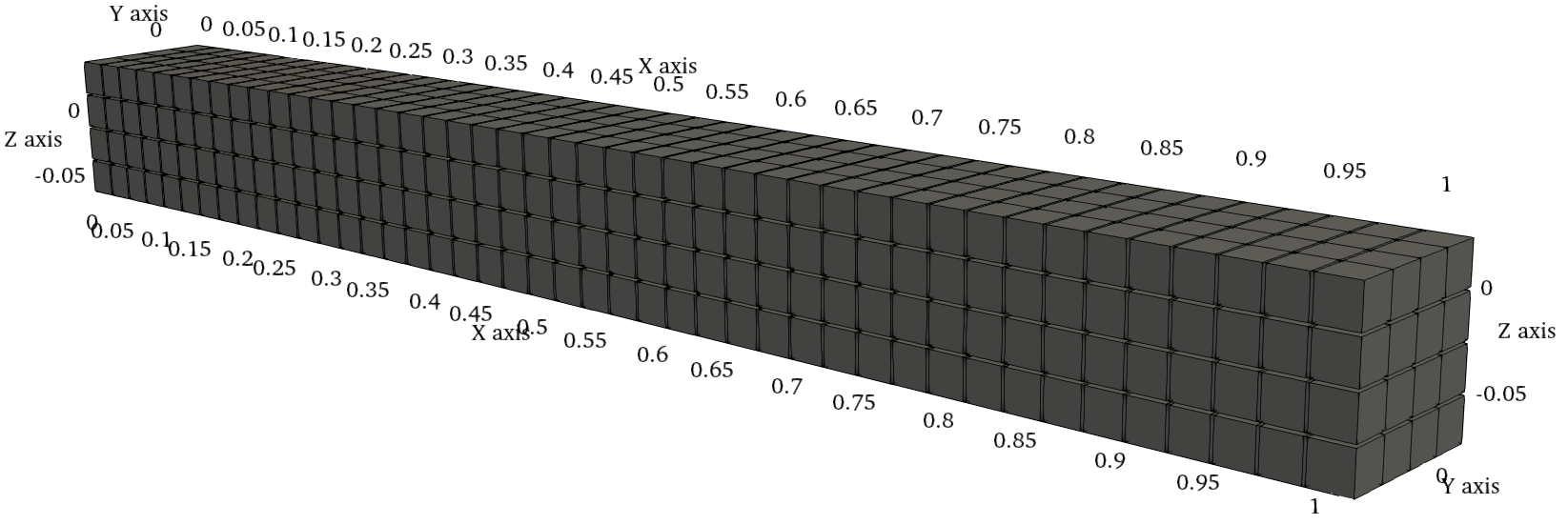}
        \caption{$t=10$}
        \label{fig:cantilevered-beam-final}
    \end{subfigure}
    \caption{A three dimensional cantilevered beam: (a) the initial condition, (b) the bonds between material points, and (c) the state given a constant external acceleration $a = -10^{-1 }$ $m/s^2$ in the $z$ direction at time $t=10$ $s$. We set the physical parameters $M=1$ $kg$ and $E=10^{3}$ $Pa$. The left column shows bonds only between adjacent material points and the right column shows bonds within a fixed radius $r=0.06$ $m$.}
    \label{fig:cantilevered-beam}
\end{figure}

The left column of Figure \ref{fig:cantilevered-beam} shows a three dimensional cantilevered beam where the material points are only bonded with their nearest neighbors. As the number of points increases, the solution neighborhood containing the connected points decreases and the numerical model recovers the analytical solution. The right column fixes a constant radius---as the number of points increases a given point connects with more nearest neighbors. The latter simulation reflects the non-local forces used by peridynamic models. Physically, this could represent internal forces that evolve on near-instantaneous time scales. Mathematically, this example is more complicated than a typical peridynamic model. In particular, we do not have an explicit expression for the force state and we also need to explicitly evolve the angular momentum conservation equation. However, the peridynamic-like model with non-local forces has the intuitive effect of stiffening the beam---as shown in Figure \ref{fig:cantilevered-beam-final}. We leave exploring the mathematics more rigorously to future work.

\section{Conclusions and future directions}

ParticLS is a software package for discrete element and peridynamic methods designed with geoscience applications in mind. Our novel optimization-based contact detection routine allows users to define complex particle geometries via an SDF. Leveraging abstraction and inheritance allows users to easily add new geometries without re-implementing contact-detection algorithms. Also, the ability to evolve general time-dependent objects allows users to infuse additional physics into their simulations beyond typical particle mechanics. ParticLS can also couple rigid body dynamics and peridynamic interactions in the same simulation, which allows users to model large bodies that fracture into smaller fragments that then interact with each other in a discrete manner. These interactions can be fine-tuned for a particular material through the definitions of new contact models or peridynamic states that capture the appropriate mechanics.

ParticLS focuses on creating computational objects that mimic mathematical counterparts (e.g. abstract classes that implement signed distance functions to represent particle geometries). Without this general framework the software must either assume a particular geometry (typically spheres) or implement different pair-wise contact laws for each possible pair of particle geometries.

In summary, ParticLS provides a platform with an intuitive implementation that will assist further study of the constitutive laws governing materials prone to fracture and plastic deformation, specifically rock and ice, as well as empower the simulation of systems that are relevant to various geoscience fields. In particular, ParticLS' ability to simultaneously approximate the peridynamic equations and evolve time-dependent parameters, such as temperature, could allow users to model temperature-dependent interaction laws. For example, materials may become brittle as temperatures decrease or ductile as temperatures increase. This capability is particularly useful when modeling the mechanical response of land or sea ice for simulating glacier calving or sea ice pressure ridging. The coupling of DEM and peridynamics is particularly useful for modeling ice ridging, where individual ice fragments coalesce to form sails and keels that catch atmospheric and oceanic currents \cite{Timco1997,Hopkins1991}. 
These examples of future model developments and applications illustrate how ParticLS provides a useful set of base functionality that can be extended and used to study a variety of geoscience problems with complicated physics.

\begin{acknowledgements}
The authors gratefully acknowledge funding provided by the Engineering Research and Development Center (ERDC) Future Innovation Funds (FIF) program. Since writing this paper, author Andrew D. Davis has moved from the Cold Regions Research and Engineering Labs to the Courant Institute at New York University, where he is funded by the Multidisciplinary University Research Initiative---ONR N00014-19-1-2421.
\end{acknowledgements}

\paragraph{Conflict of Interest}

On behalf of all authors, the corresponding author states that there is no conflict of interest.



\end{document}